\documentclass[aps,prl,showkeys,showpacs,
reprint,
floatfix]{revtex4-1}

\usepackage[colorlinks=true,hyperfootnotes=true,breaklinks=true]{hyperref}
\usepackage[utf8]{inputenc}
\usepackage[T1]{fontenc}
\usepackage{amsmath}
\usepackage{graphicx}
\usepackage[figtopcap]{subfigure}
\usepackage{psfrag}
\usepackage{auto-pst-pdf}

\usepackage{amssymb}
\usepackage[finalnew]{trackchanges}
\newcommand{\xplus}{%
  \ensuremath{\vcenter{\offinterlineskip\hbox{$+$}\vskip-1.55ex\hbox{$\times$}}}
}

\addeditor{Rev.1} 
\addeditor{Rev.2} 

\begin{document}
\title{Simple cubic random-site percolation thresholds \change[Rev.1]{beyond Rubik's}{for }neighborhoods\change[Rev.2]{ with next-next-next-nearest neighbors}{ containing fourth-nearest neighbors}}
\author{Krzysztof Malarz}
\homepage{http://home.agh.edu.pl/malarz/}
\email{malarz@agh.edu.pl}
\affiliation{\href{http://www.agh.edu.pl/}{AGH University of Science and Technology},
\href{http://www.pacs.agh.edu.pl/}{Faculty of Physics and Applied Computer Science},
al. Mickiewicza 30, 30-059 Krakow, Poland.}

\begin{abstract}
In the paper random-site percolation thresholds for simple cubic lattice
with sites' neighborhoods containing next-next-next-nearest
neighbors (4NN) are evaluated with Monte Carlo simulations.
A recently proposed algorithm with low sampling for percolation
thresholds estimation [Bastas {\em et al.}, arXiv:1411.5834] is
implemented for the studies of the top-bottom 
wrapping probability. The obtained percolation thresholds are
$p_C(\text{4NN})=0.3116$\add{0}$(1$\add{2}$)$, $p_C(\text{4NN+NN})=0.1504$\add{0}$(1$\add{2}$)$,
$p_C(\text{4NN+2NN})=0.1595$\add{0}$(1$\add{2}$)$, $p_C(\text{4NN+3NN})=0.2049$\add{0}$(1$\add{2}$)$,
$p_C(\text{4NN+2NN+NN})=0.1144$\add{0}$(1$\add{2}$)$,
$p_C(\text{4NN+3NN+NN})=0.1192$\add{0}$(1$\add{2}$)$,
$p_C(\text{4NN+3NN+2NN})=0.1133$\add{0}$(1$\add{2}$)$,
$p_C(\text{4NN+3NN+2NN+NN})=0.1000$\add{0}$(1$\add{2}$)$,
where 3NN, 2NN, NN stands for next-next-nearest neighbors, next-nearest neighbors, and nearest neighbors, respectively. 
\add{As an SC lattice with 4NN neighbors may be mapped onto two independent interpenetrated SC lattices but with two times larger lattice constant the percolation threshold $p_C$(4NN) is exactly equal to $p_C$(NN).}
\add{The simplified Bastas \emph{et al.} method allows for reaching uncertainty of the percolation threshold value $p_C$ similar to those obtained with classical method but ten times faster.}
\end{abstract}



\date{\today}
\maketitle

%% ##########################################################
\section{Introduction}
%% ##########################################################

Finding percolation thresholds $p_C$ and observing cluster properties near percolation threshold \cite{bookDS,bookBB,bookHK,bookMS} are one of the
most extensively studied problems in statistical physics.
The beauty of percolation \cite{Broadbent1957,*Frisch1961,*Frisch1962}
lays both in its simplicity and possible practical applications.
The latter ranges from theoretical studies of geometrical model of the
phase transition
\cite{PhysRevE.90.012815,*Wierman2005,*Rosowsky2000,*ISI:A1983QF02000016,*ISI:A1976BD65000014,*Sykes1976b,*Sykes1976c,*Sykes1976d,*Gaunt1976},
via condensed matter physics
\cite{PhysRevB.89.054409,*Silva2011,*Shearing2010,*Halperin2010},
rheology \cite{Mun2014,*Amiaz2011,*Bolandtaba2011,*Mourzenko2011},
forest fires
\cite{Abades2014,*Camelo-Neto2011,*Guisoni2011,*Simeoni2011,*Kaczanowska2002}
to immunology
\cite{Silverberg2014,*Suzuki2011,*Lindquist2011,*Naumova2008,*Floyd2008}
and quantum mechanics \cite{Chandrashekar2014}.

In random-site percolation model the nodes of lattice, graph or network are randomly occupied with a probability $p$.
The critical probability $p_C$ separates two phases:
for $p>p_C$ the system percolates, {\em i.e.} one may find a single 
 cluster of occupied sites which extends to the borders of
the system; while for $p<p_C$ only smaller clusters exist.
Usually, the finite size scaling theory
\cite{Fisher1971,bookVP,Binder1992,bookDL} is employed for
percolation threshold $p_C$ estimation. This requires checking
properties of some quantity $X(p,L)$ in the vicinity of phase
transition as it depends on the linear system size $L$
\begin{equation}
\label{eq-scaling}
X(p;L) = L^{-x}\cdot\mathcal{F}\left((p-p_C)L^{1/\nu}\right),
\end{equation}
where $\mathcal{F}(\cdot)$ is a scaling function, $x$ is a scaling exponent and $\nu$ is a critical exponent associated with the correlation length \cite{bookDS}.
Eq.~\eqref{eq-scaling} yields an efficient way for $p_C$ determination as $L^x\cdot X(p_C;L)=\mathcal{F}(0)$ does not depend on the linear system size $L$.
It means that curves $L^x\cdot X(p;L)$ plotted for various values of $L$ should have one common point exactly at $p=p_C$.
Unfortunately, the results of computer simulations rather rarely reproduce a 
single common point of curves $X(p;L)$
unless the number $N_{\text{run}}$ of prepared lattices is very high.

Recently, Bastas {\em et al.} proposed efficient method for estimating
scaling exponents $x$ and percolation thresholds $p_C$ in
percolation processes with low
sampling~\cite{PhysRevE.84.066112,Bastas2014}.
According to Refs.~\cite{Bastas2014,PhysRevE.84.066112} instead of searching for
the point where curves $X(p;L)$ intercept each other one may wish to
minimize the pairwise difference
\begin{equation}
\label{eq-Lambda}
\Lambda(p;x)\equiv\sum_{i\ne j}\left[H(p;L_i)-H(p;L_j)\right]^2,
\end{equation}
with respect to both parameters $x$ and $p$, where
\begin{subequations}
\label{eq-H}
\begin{equation}
H(p;L)\equiv Y(p;L)
\end{equation}
as suggested in Ref.~\cite{PhysRevE.84.066112} or
\begin{equation}
H(p;L)\equiv Y(p;L) + 1/Y(p;L)
\end{equation}
\end{subequations}
as proposed in Ref.~\cite{Bastas2014} and in both cases
\begin{equation}
\label{eq-Y}
Y(p;L)\equiv L^x \cdot X(p;L).
\end{equation}
The minimum of $\Lambda(p;x)$ is reached for $p=p_C$ and $x=\beta/\nu$, where $\beta$ is a critical exponent associated with the order parameter (for instance  probability of an arbitrary site belonging to the infinite cluster \cite{bookDS}).

In this paper we propose simplified version of Bastas {\em et al.} algorithm,
where only a single-parameter function $\lambda(p)$ must be minimized in order to provide
percolation threshold estimation. With such approach we estimate
simple cubic (SC) random-site percolation thresholds for eight
complex neighborhoods containing next-next-next-nearest neighbors.
Our results enhance those of the earlier studies regarding percolation thresholds for complex neighborhoods on square \cite{Galam2005a,*Galam2005b,*Majewski2007} or SC \cite{Kurzawski2012} lattices.

%% ----------------------------------------------------------
\begin{figure}[!Htp]
\psfrag{2N}{NN}
\psfrag{3N}{2NN}
\psfrag{4N}{3NN}
\psfrag{5N}{4NN}
\includegraphics[width=0.5\columnwidth]{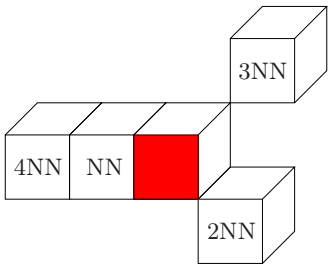}
\caption{\label{fig-sites} Single sites from various neighborhoods of
an %%
SC lattice. The full neighborhoods contain $z=6$, 12, 8 and 6
sites for NN, 2NN, 3NN and 4NN neighborhoods, respectively.}
\end{figure}
%% ----------------------------------------------------------

%% ##########################################################
\section{\label{sec-approach} Approach}
%% ##########################################################

Our proposition is to apply Bastas {\em et al.} technique for such quantity $X(p;L)$ which does not require scaling along $X$ axis by a factor $L^x$
in order to achieve
statistical invariance of the shape $X(p;L)$ for various values of
$L$. An example of such quantity is the (top-bottom) wrapping probability \footnote{The detailed studies of properties $W(p;L)$ and its scaling function $\mathcal{G}(\cdot)$ but for a square lattice are given in Ref.~\cite{Newman2001}.}:
\begin{equation}
\label{eq-W}
W(p;L)={N(p;L)}/{N_{\text{run}}},
\end{equation}
where $N(p;L)$ is a number of percolating lattices with $p\cdot L^3$ occupied sites among $N_{\text{run}}$ lattices constructed for fixed values $p$ and $L$.
In thermodynamic limit we have $W(p<p_C;L\to\infty)=0$ and $W(p>p_C,L\to\infty)=1$ and thus scaling exponent $x$ of $W$ is
equal to zero \cite{bookDS}.
Consequently, instead of the form given by Eq.~\eqref{eq-scaling} wrapping probability obeys a simplified scaling relation \cite{bookDS,Newman2001}
\begin{equation}
\label{eq-WG}
W(p;L) = \mathcal{G}\left((p-p_C)L^{1/\nu}\right).
\end{equation}
Eq.~\eqref{eq-WG} again makes possible $p_C$ determination as $W(p_C;L)=\mathcal{G}(0)$ does not depend on the system size $L$.

Now, the equivalent of Eq.~\eqref{eq-Lambda} may be written as
\begin{equation}
\label{eq-lambda}
\lambda(p)\equiv\sum_{i\ne j}\left[H(p;L_i)-H(p;L_j)\right]^2,
\end{equation}
where
\begin{equation}
H(p;L)\equiv W(p;L) + 1/W(p;L).
\end{equation}
Following Bastas {\em et al.} technique one should minimize function $\lambda(p)$; the found minimum may be then used for the $p_C$ estimation.

Several numerical techniques allows for clusters of connected sites identification \cite{Hoshen1976a,Leath1976,Newman2001,Torin2014}.
Here we apply Hoshen--Kopelman algorithm \cite{Hoshen1976a}, which allows for sites labeling in a such way, that occupied sites in the same cluster have assigned the same labels and different clusters have different labels associated with them.

Here we investigate 
an 
SC lattice with sites' neighbors ranging from the nearest
neighbors (NN), via the next-nearest neighbors (2NN) and the
next-next-next-nearest neighbors (3NN) to the
next-next-next-next-nearest neighbors (4NN).
A scheme showing only single sites of each of the neighborhood
types mentioned above is presented in Fig.~\ref{fig-sites}. The full neighborhoods contain $z=6$, 12, 8 and
6 sites for NN, 2NN, 3NN and 4NN neighborhoods, respectively. Also
all available combinations of these neighborhoods are considered,
{\em i.e.} (4NN+NN), (4NN+2NN), (4NN+3NN), (4NN+2NN+NN),
(4NN+3NN+NN), (4NN+3NN+2NN) and (4NN+3NN+2NN+NN) containing
$z=12$, 18, 14, 24, 20, 26 and 32 sites, respectively.

%% ##########################################################
\section{\label{sec-results}Results and discussion}
%% ##########################################################

%% ----------------------------------------------------------
\begin{table}[!htbp]
\caption{\label{tab-PT} The critical values of $p_C$ for various neighborhoods based on minimization of $\lambda(p)$ function.}
\begin{ruledtabular}
\begin{tabular}{lll}
 neighborhood&$z$& $p_C$    \\ \hline
 4NN            & 6& $0.3116$\add{0}$(1$\add{2}$)=p_C(\text{NN})$ \\
 4NN+NN         &12& $0.1504$\add{0}$(1$\add{2}$)$ \\
 4NN+2NN        &18& $0.1595$\add{0}$(1$\add{2}$)$ \\
 4NN+3NN        &14& $0.2049$\add{0}$(1$\add{2}$)$ \\
 4NN+2NN+NN     &24& $0.1144$\add{0}$(1$\add{2}$)$ \\
 4NN+3NN+NN     &20& $0.1192$\add{0}$(1$\add{2}$)$ \\
 4NN+3NN+2NN    &26& $0.1133$\add{0}$(1$\add{2}$)$ \\
 4NN+3NN+2NN+NN &32& $0.1000$\add{0}$(1$\add{2}$)$ \\
\end{tabular}
\end{ruledtabular}
\end{table}
%% ----------------------------------------------------------
%% ----------------------------------------------------------
\begin{figure*}[!htbp]
\psfrag{L=}{$L=$}
\psfrag{p}{$p$}
\psfrag{PI}{$W(p;L)$}
\subfigure[4NN+3NN+2NN+NN\label{fig-2N3N4N5N}]{\includegraphics[width=0.23\textwidth]{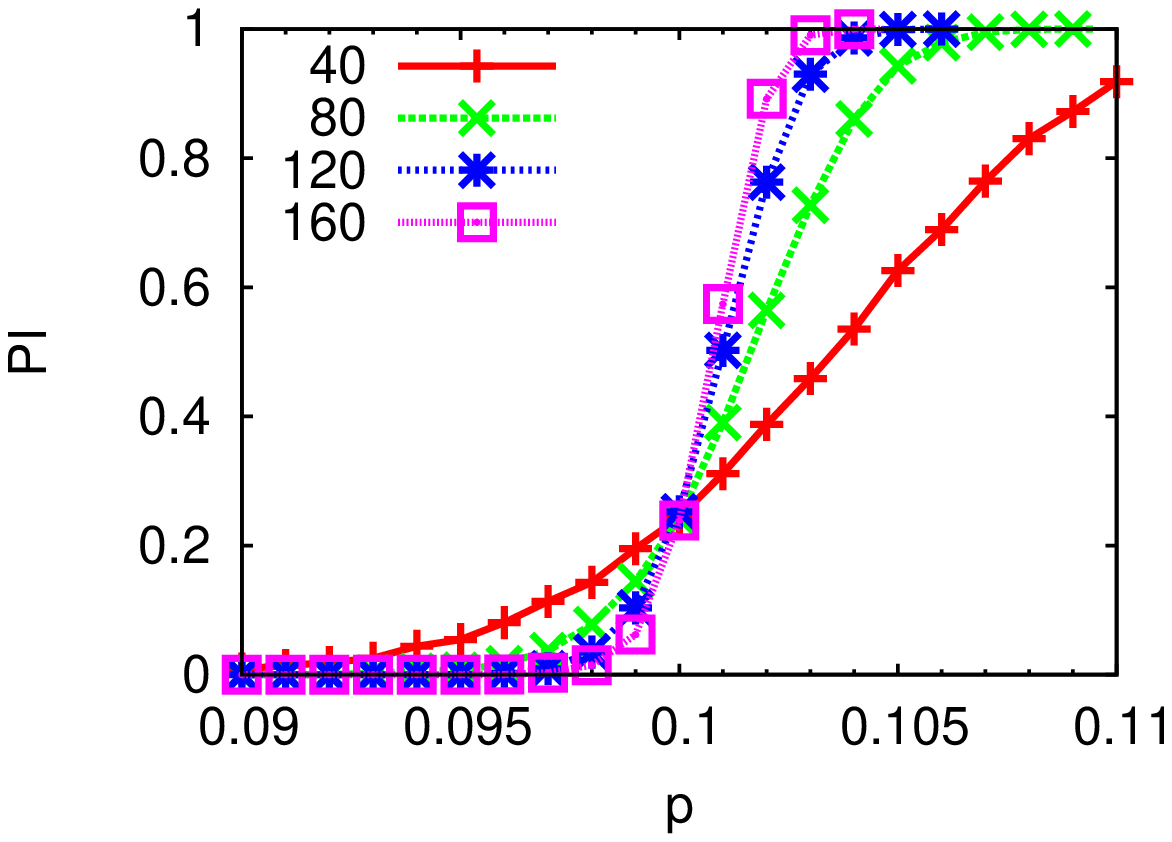}}
\subfigure[4NN+3NN+2NN\label{fig-3N4N5N}]{\includegraphics[width=0.23\textwidth]{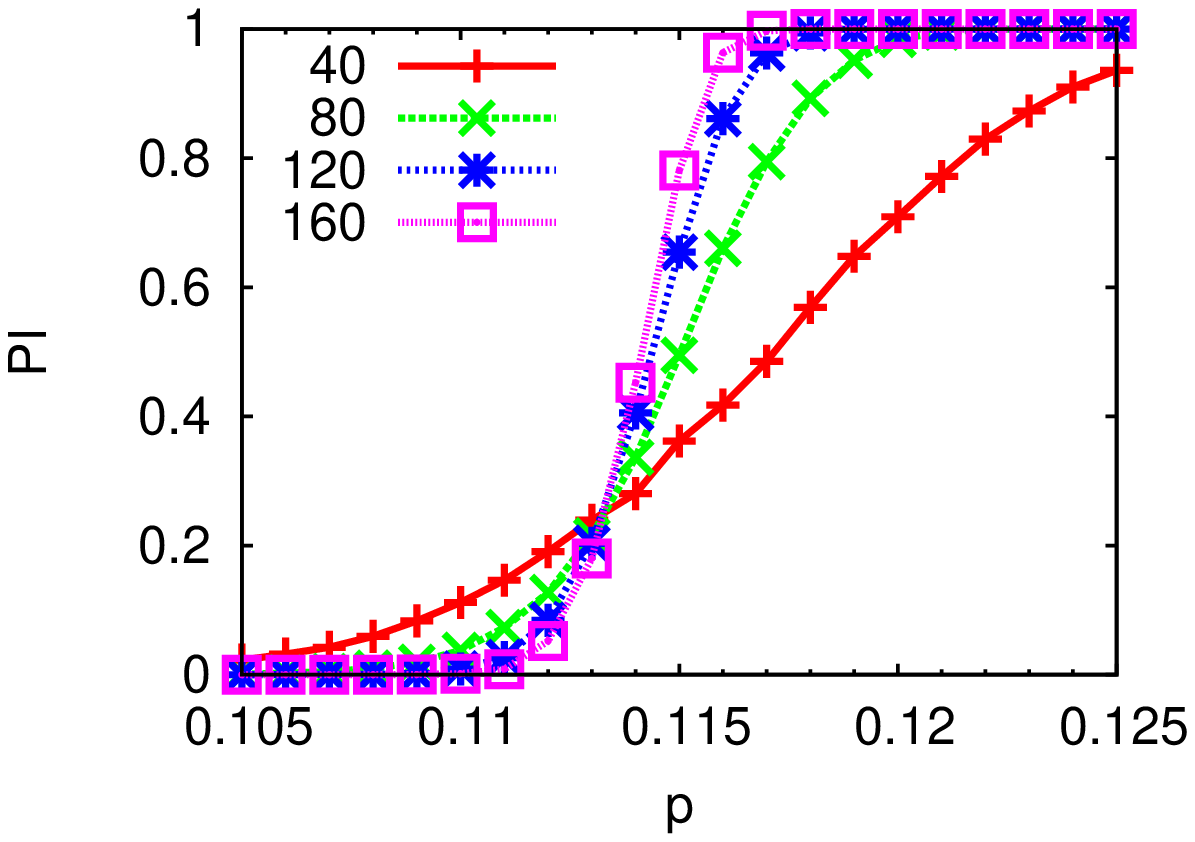}}
\subfigure[4NN+3NN+NN\label{fig-2N4N5N}]{\includegraphics[width=0.23\textwidth]{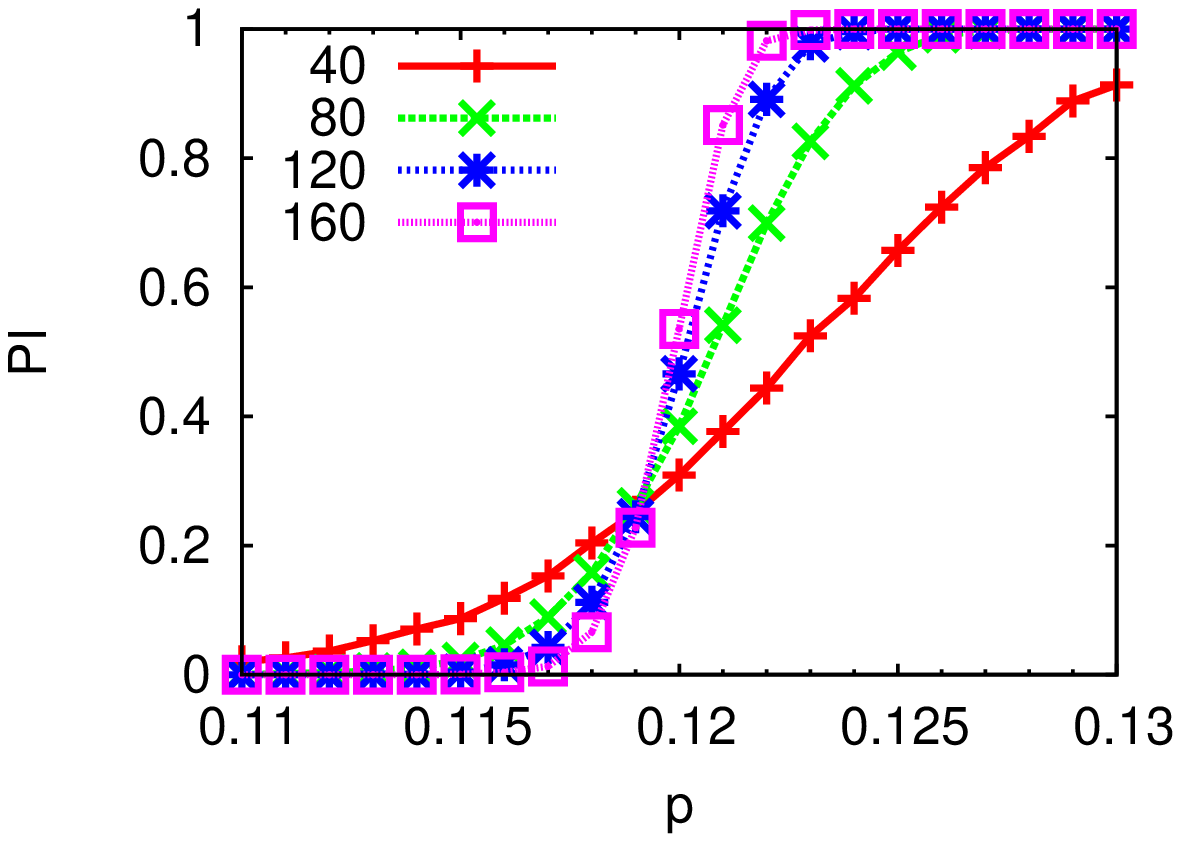}}
\subfigure[4NN+2NN+NN\label{fig-2N3N5N}]{\includegraphics[width=0.23\textwidth]{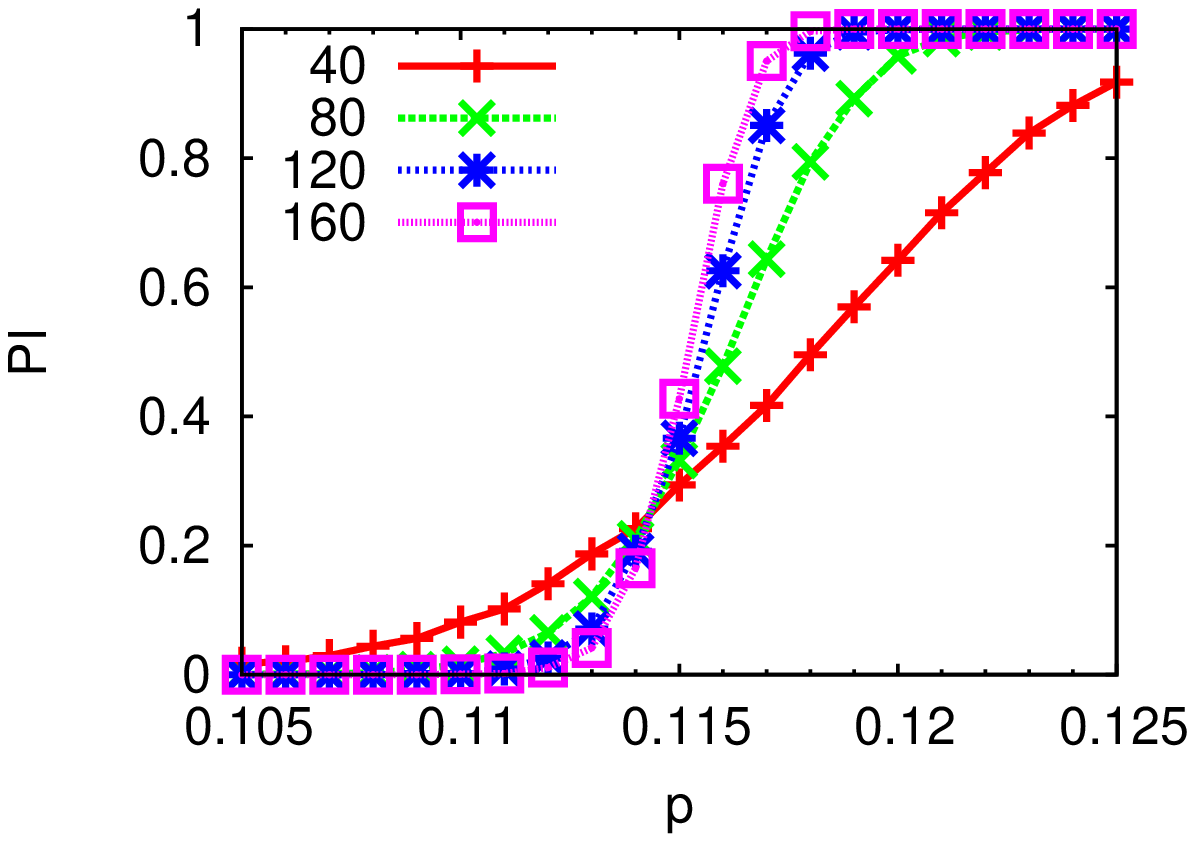}}\\
\subfigure[4NN+3NN\label{fig-4N5N}]{\includegraphics[width=0.23\textwidth]{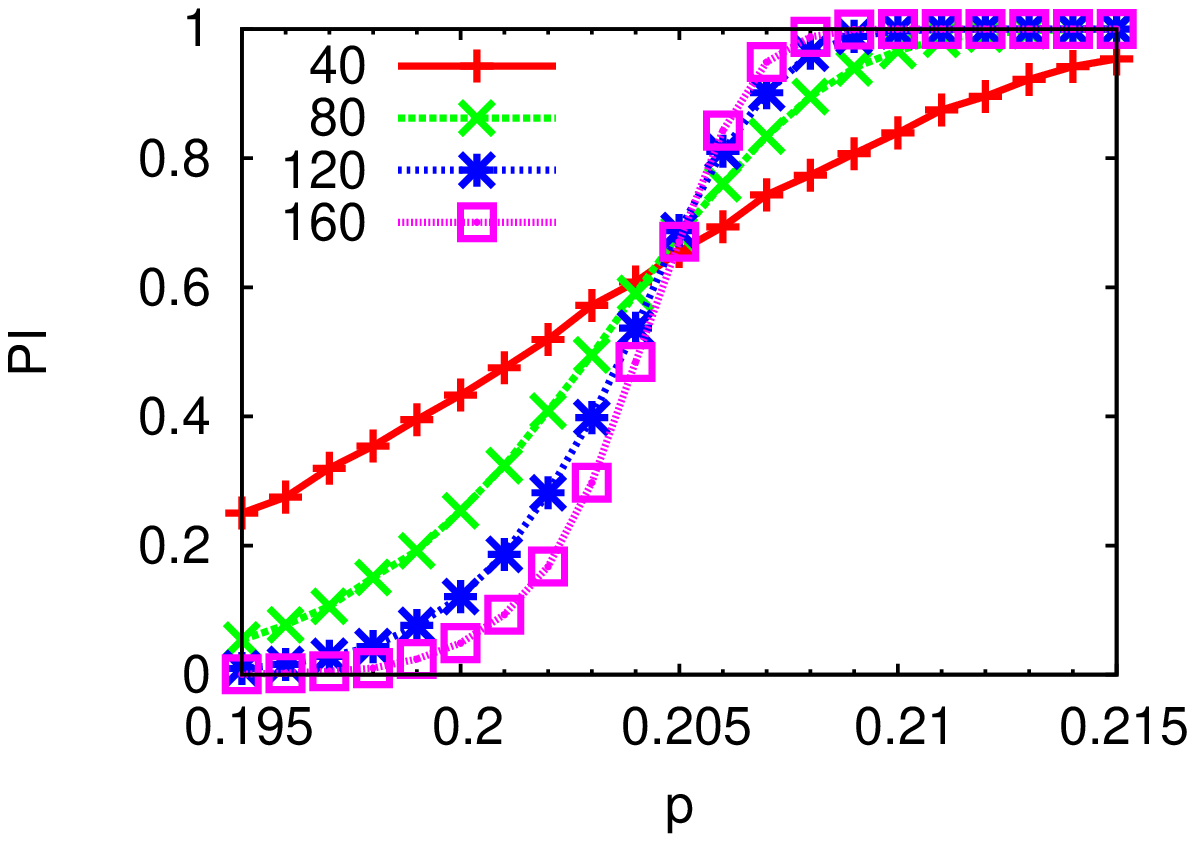}}
\subfigure[4NN+2NN\label{fig-3N5N}]{\includegraphics[width=0.23\textwidth]{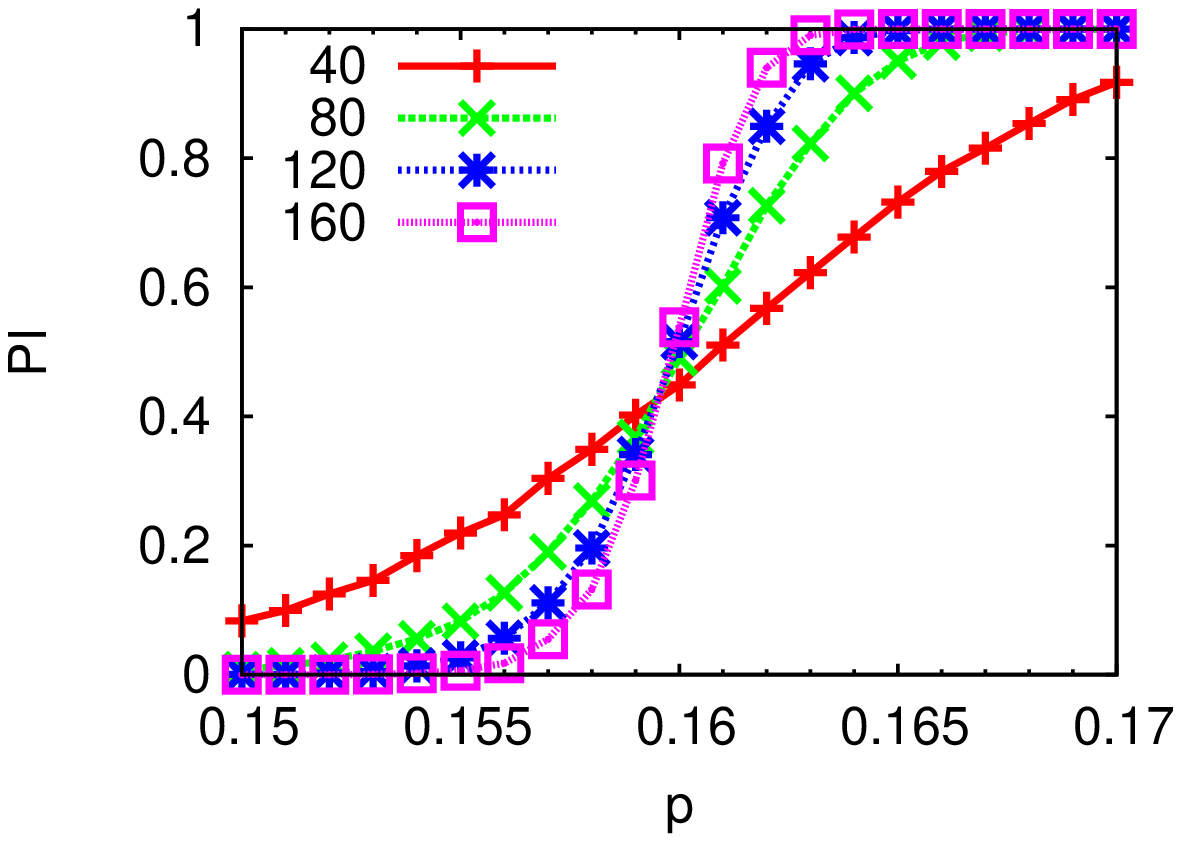}}
\subfigure[4NN+NN\label{fig-2N5N}]{\includegraphics[width=0.23\textwidth]{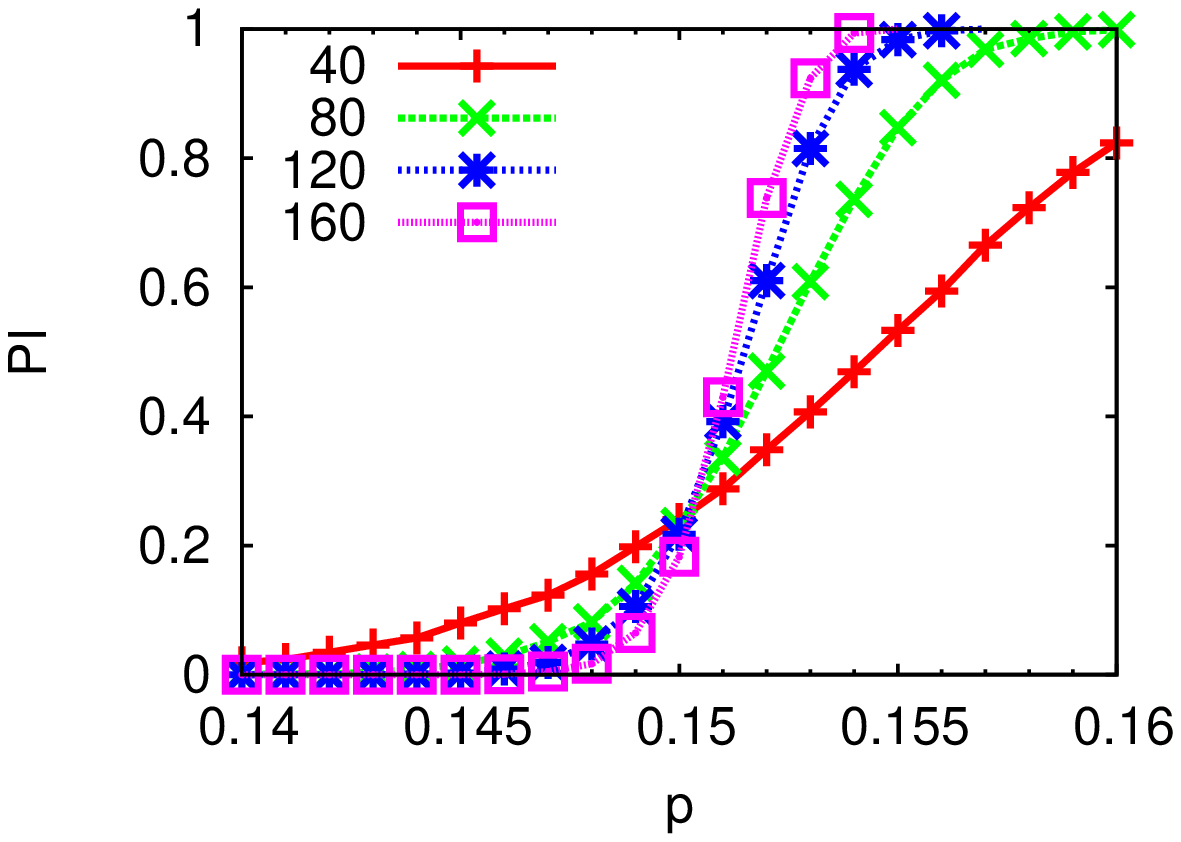}}
\subfigure[4NN\label{fig-5N}]{\includegraphics[width=0.23\textwidth]{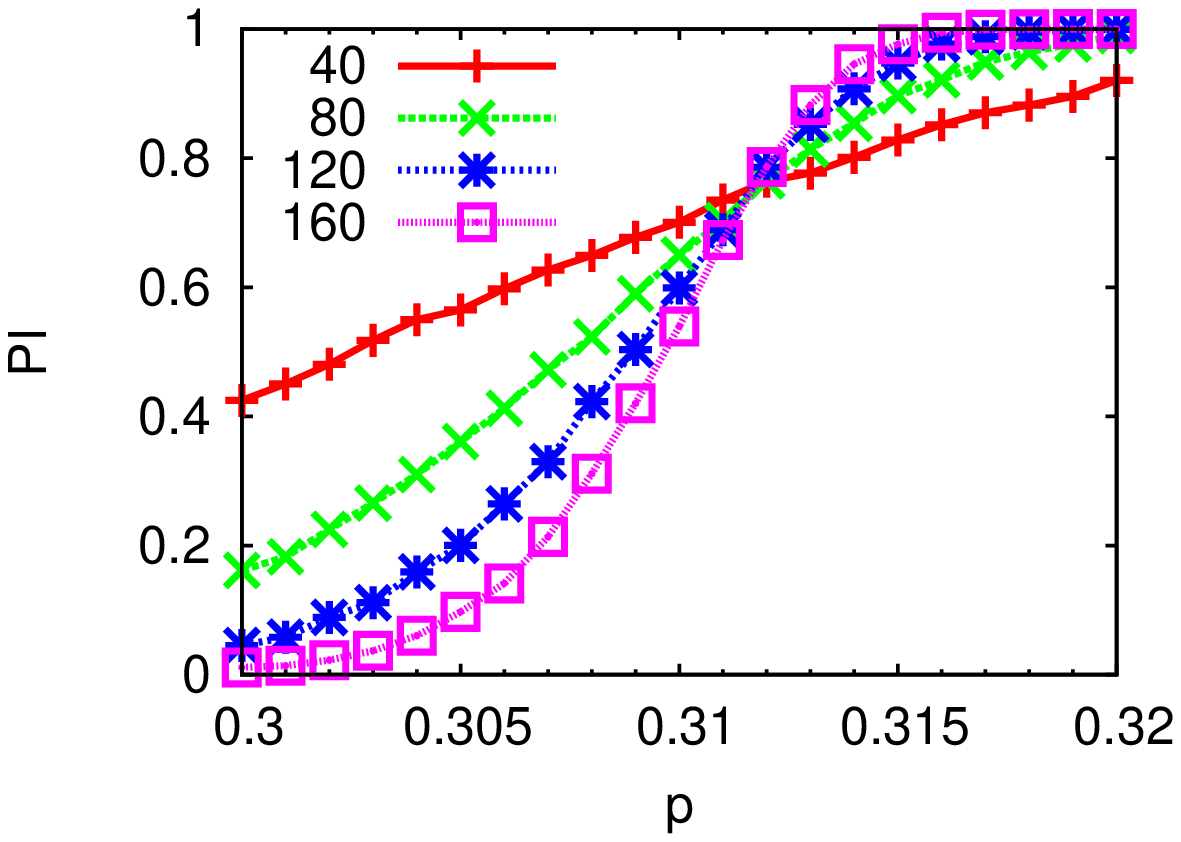}}
\caption{\label{fig-W} \add[Rev.2]{(Color online).}
Wrapping probability $W(p;L)$ vs. occupation probability $p$.
The results are averaged over $N_{\text{run}}=10^4$ runs.
The \change[Rev.2]{line colors (red, green, blue, violet)}{symbols ($+$, $\times$, $\xplus$, $\square$)} indicate the system linear sizes ($L=40$, 80, 120, 160), respectively.}
\end{figure*}
%% ----------------------------------------------------------
%% ----------------------------------------------------------
\begin{figure*}[!htbp]
\psfrag{p}{$p$}
\psfrag{L}{$\lambda(p)$}
\psfrag{PI}{$W(p;L)$}
\subfigure{\includegraphics[width=0.23\textwidth]{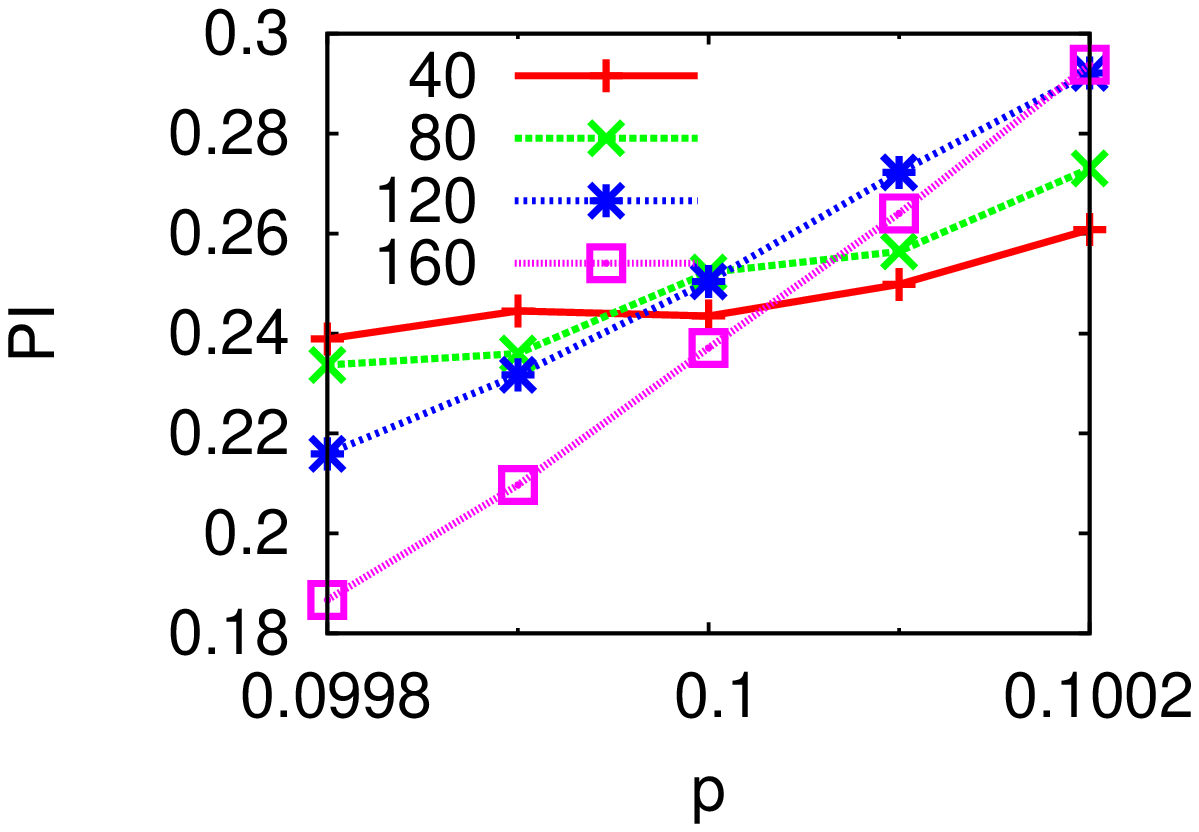}}
\subfigure{\includegraphics[width=0.23\textwidth]{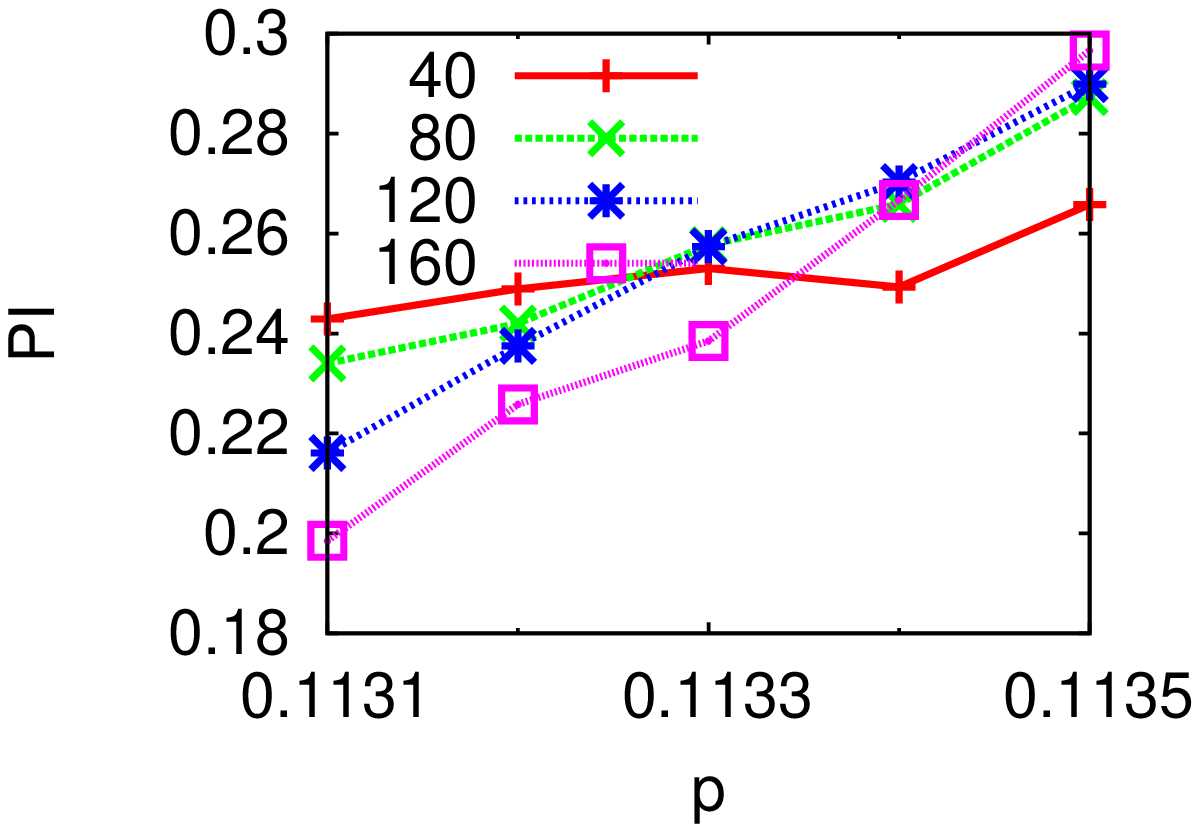}}
\subfigure{\includegraphics[width=0.23\textwidth]{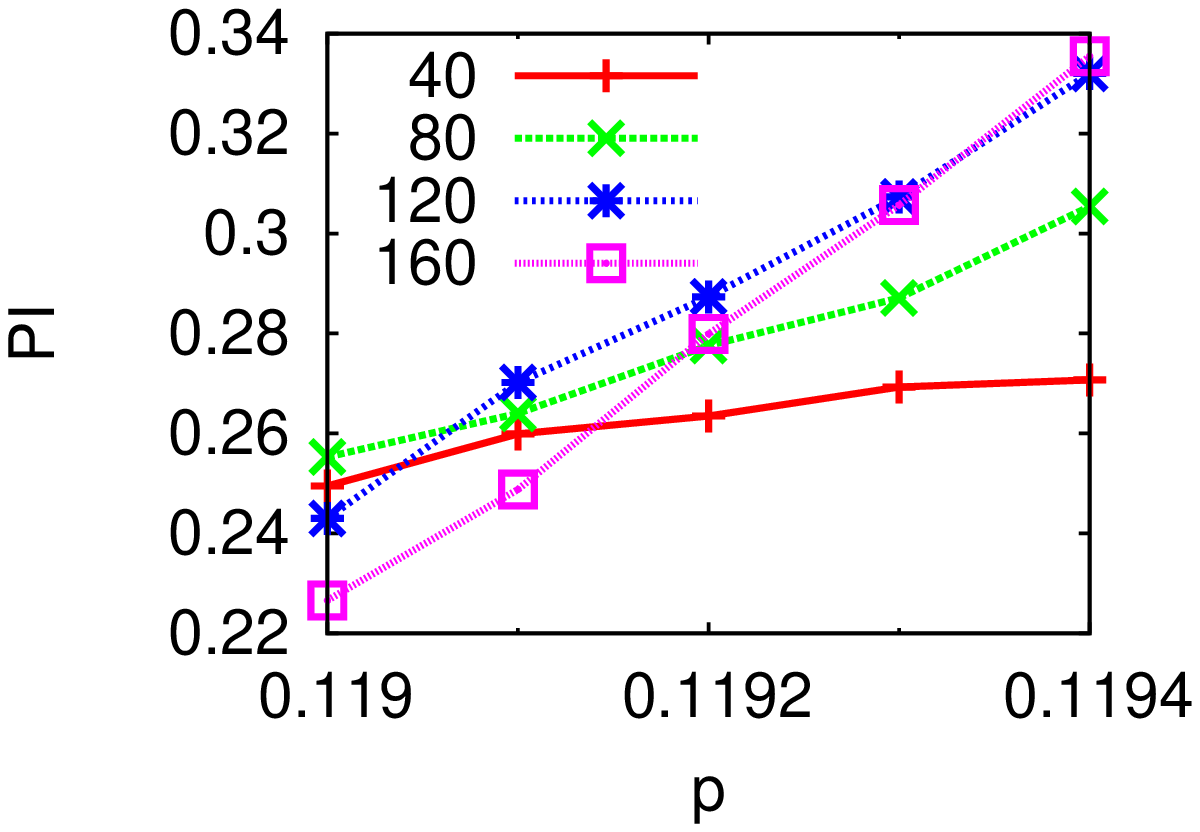}}
\subfigure{\includegraphics[width=0.23\textwidth]{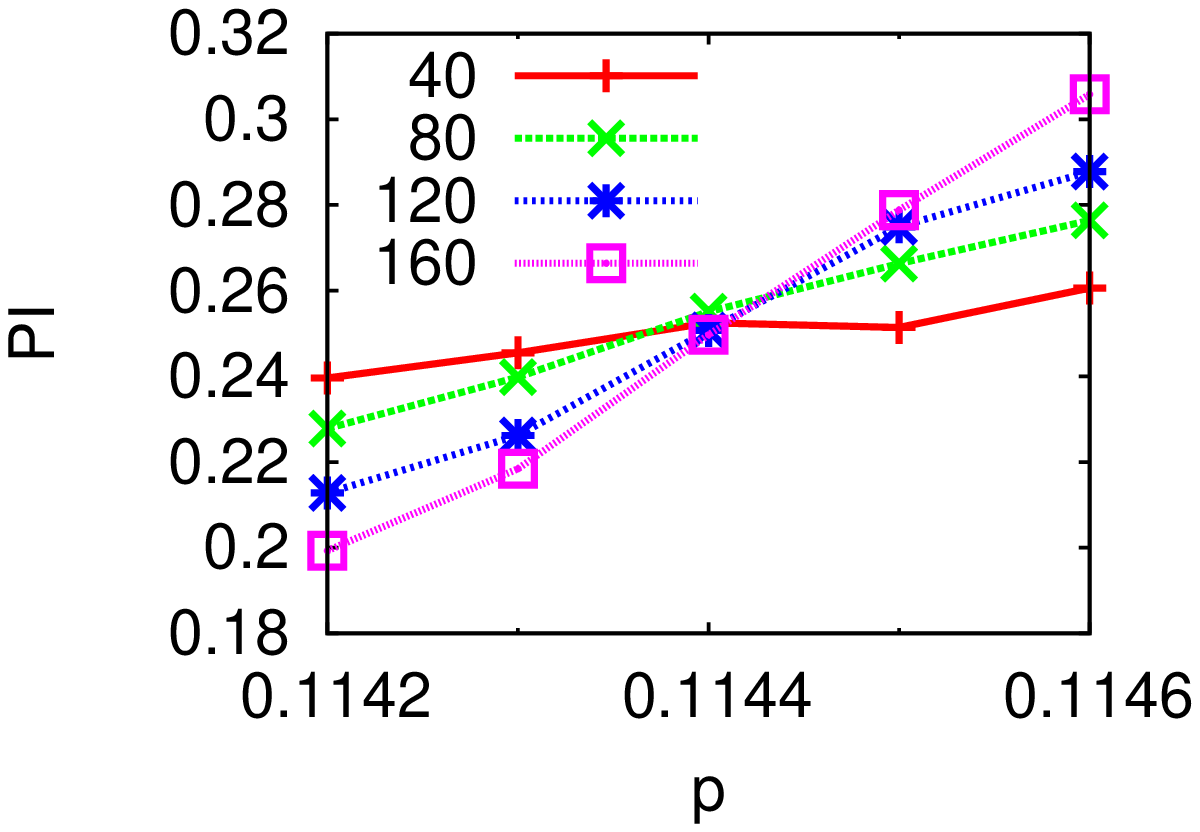}}\\
\addtocounter{subfigure}{-4}
\subfigure[5NN+4NN+3NN+2NN\label{fig-Lambda2N3N4N5N}]{\includegraphics[width=0.23\textwidth]{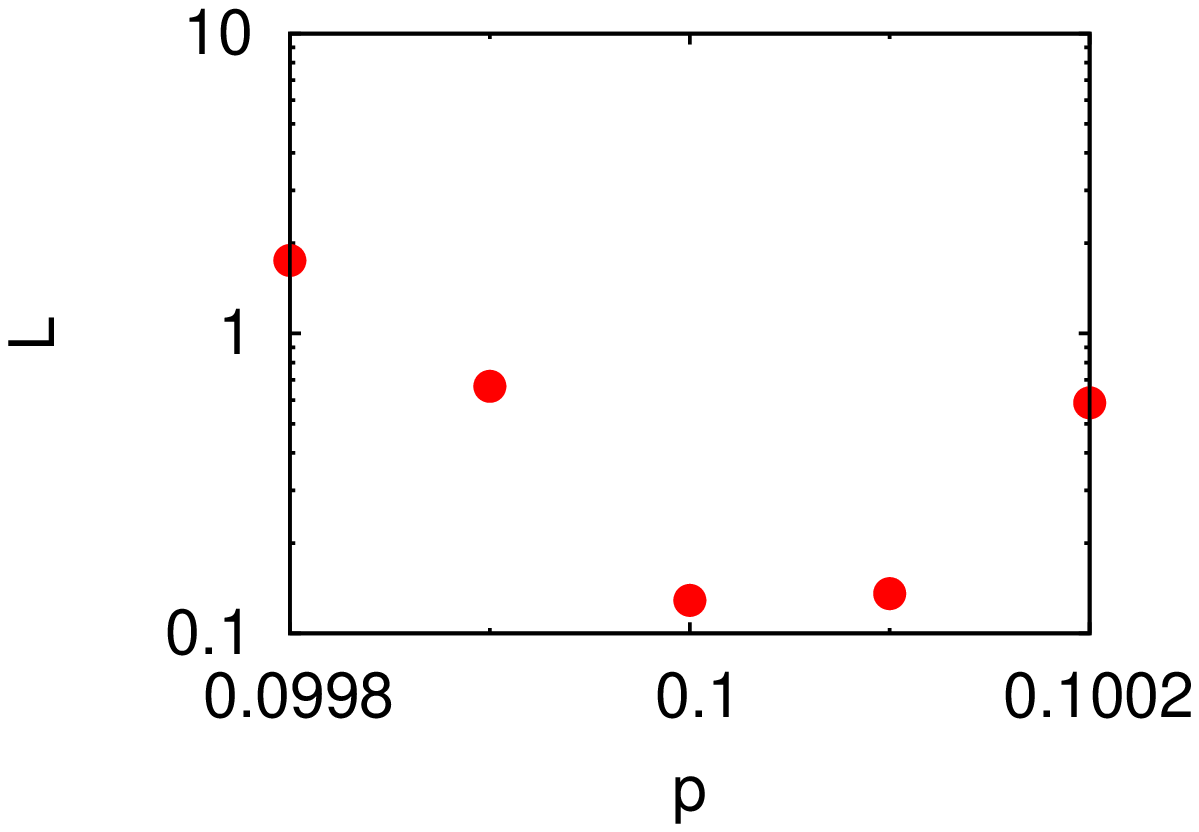}}
\subfigure[4NN+3NN+2NN\label{fig-Lambda3N4N5N}]{\includegraphics[width=0.23\textwidth]{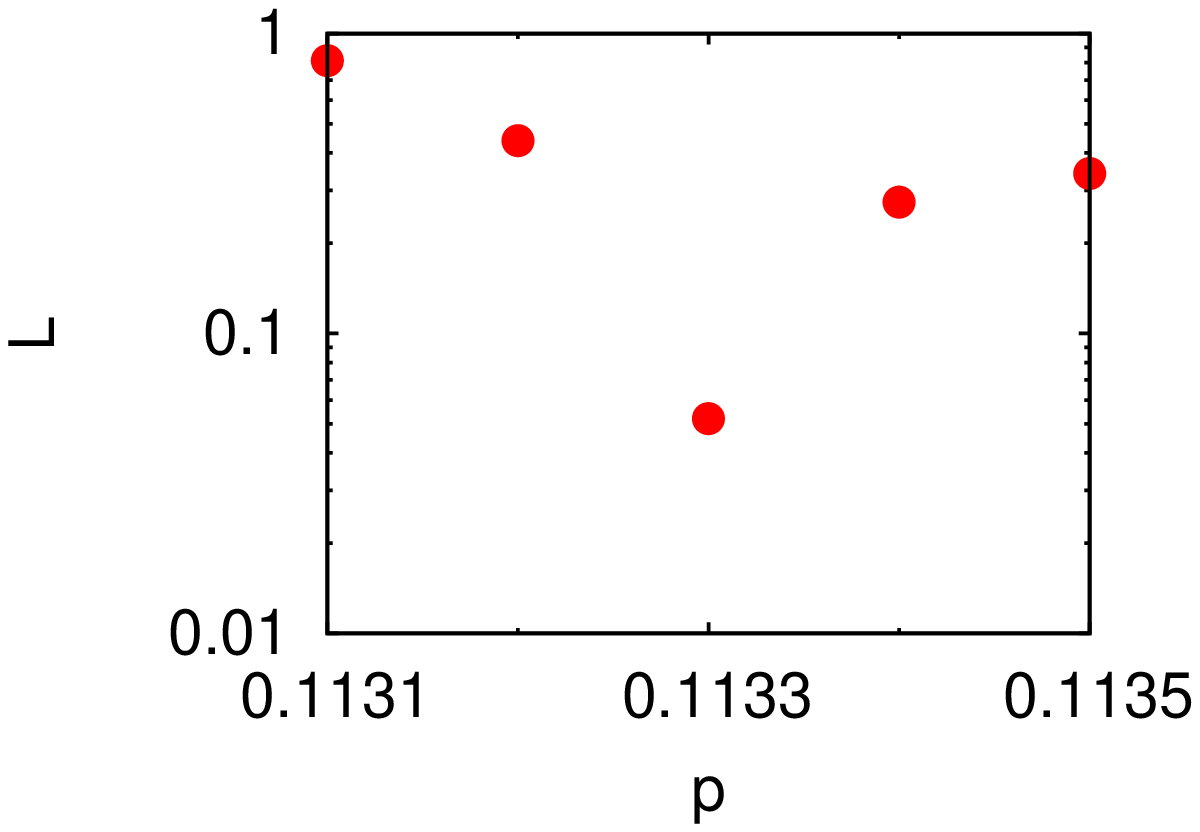}}
\subfigure[4NN+3NN+NN\label{fig-Lambda2N4N5N}]{\includegraphics[width=0.23\textwidth]{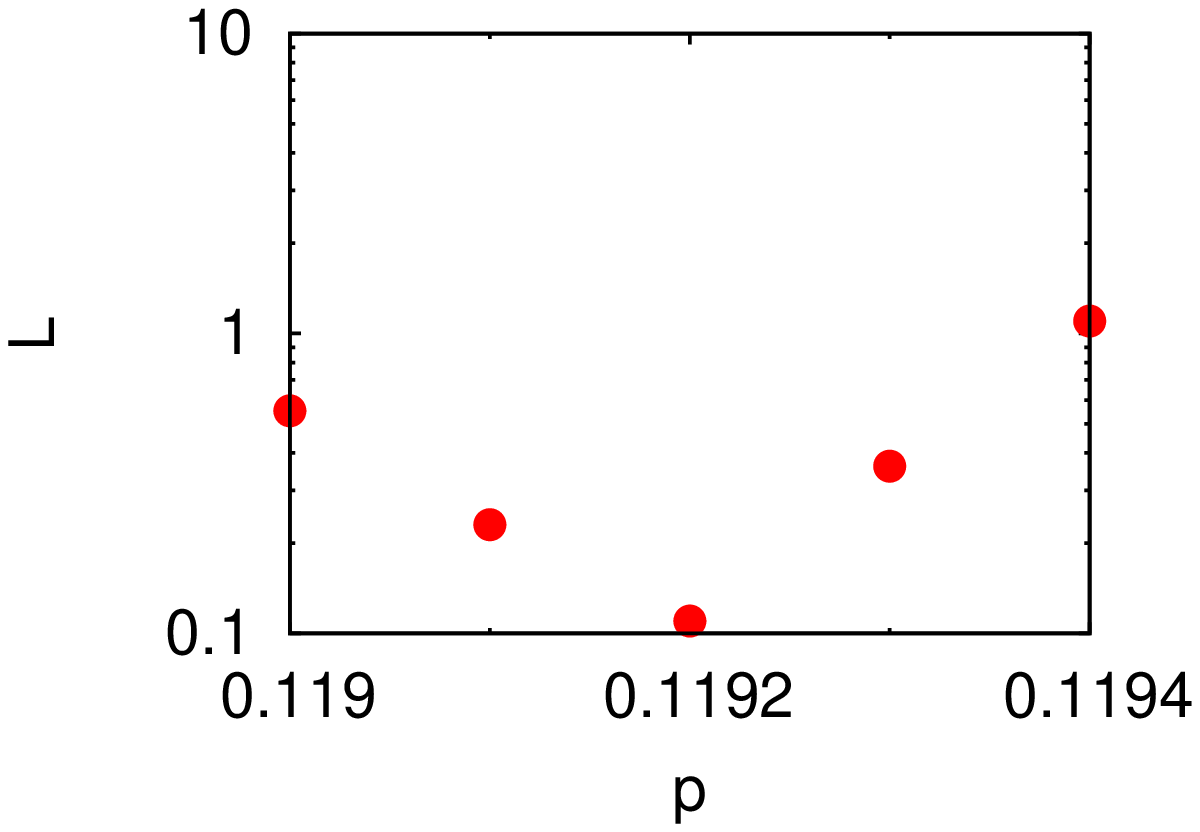}}
\subfigure[4NN+2NN+NN\label{fig-Lambda2N3N5N}]{\includegraphics[width=0.23\textwidth]{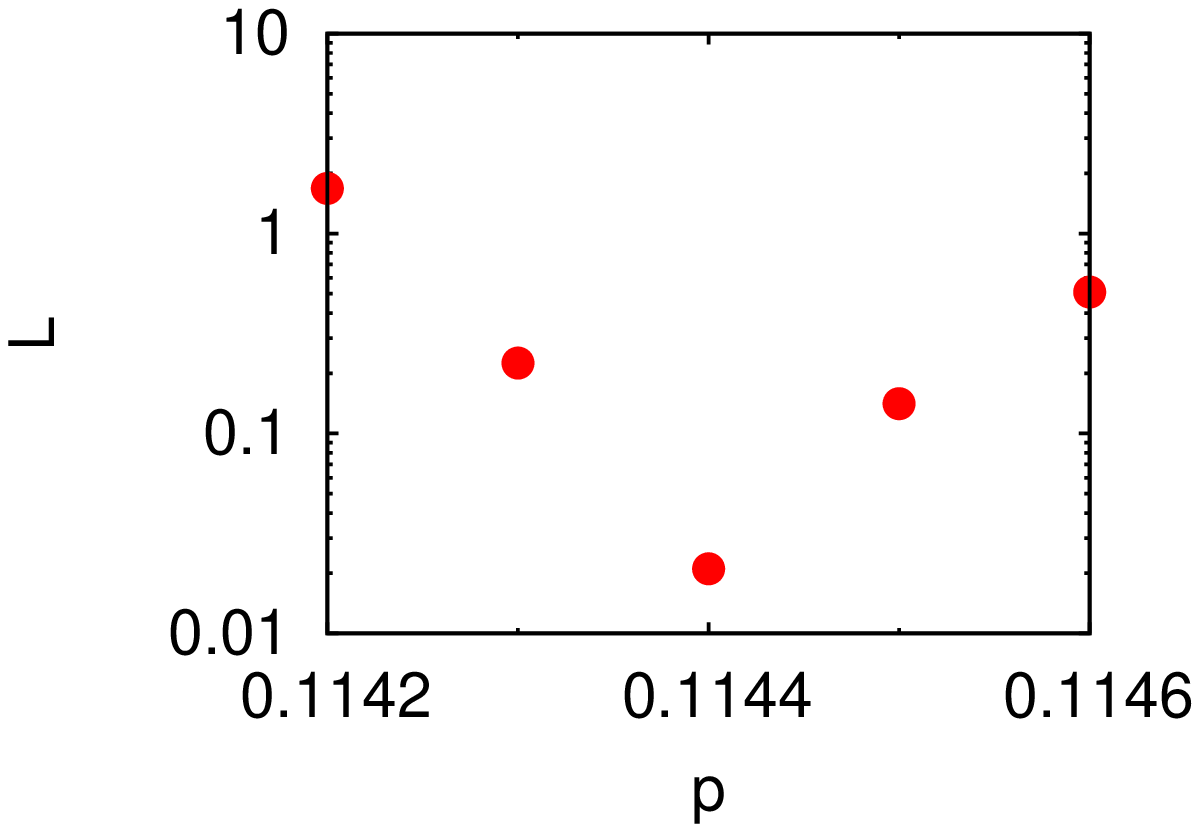}}\\
\subfigure{\includegraphics[width=0.23\textwidth]{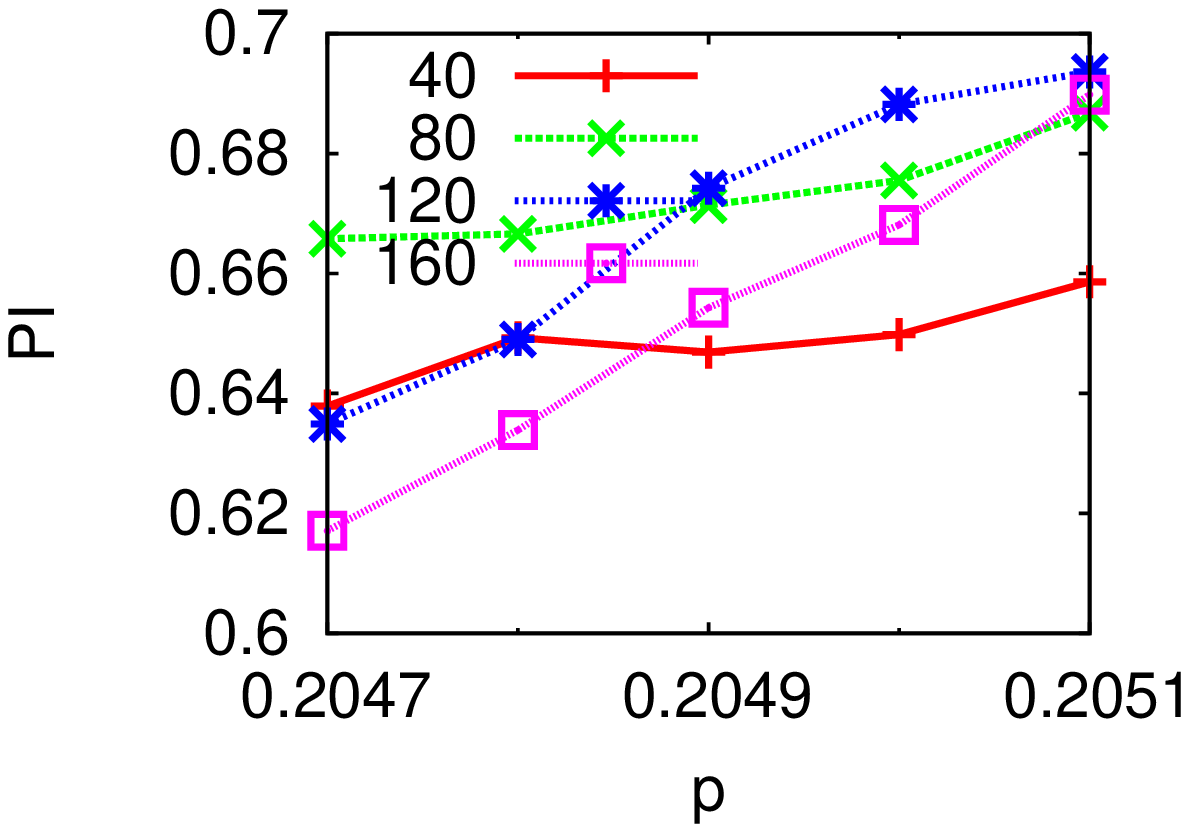}}
\subfigure{\includegraphics[width=0.23\textwidth]{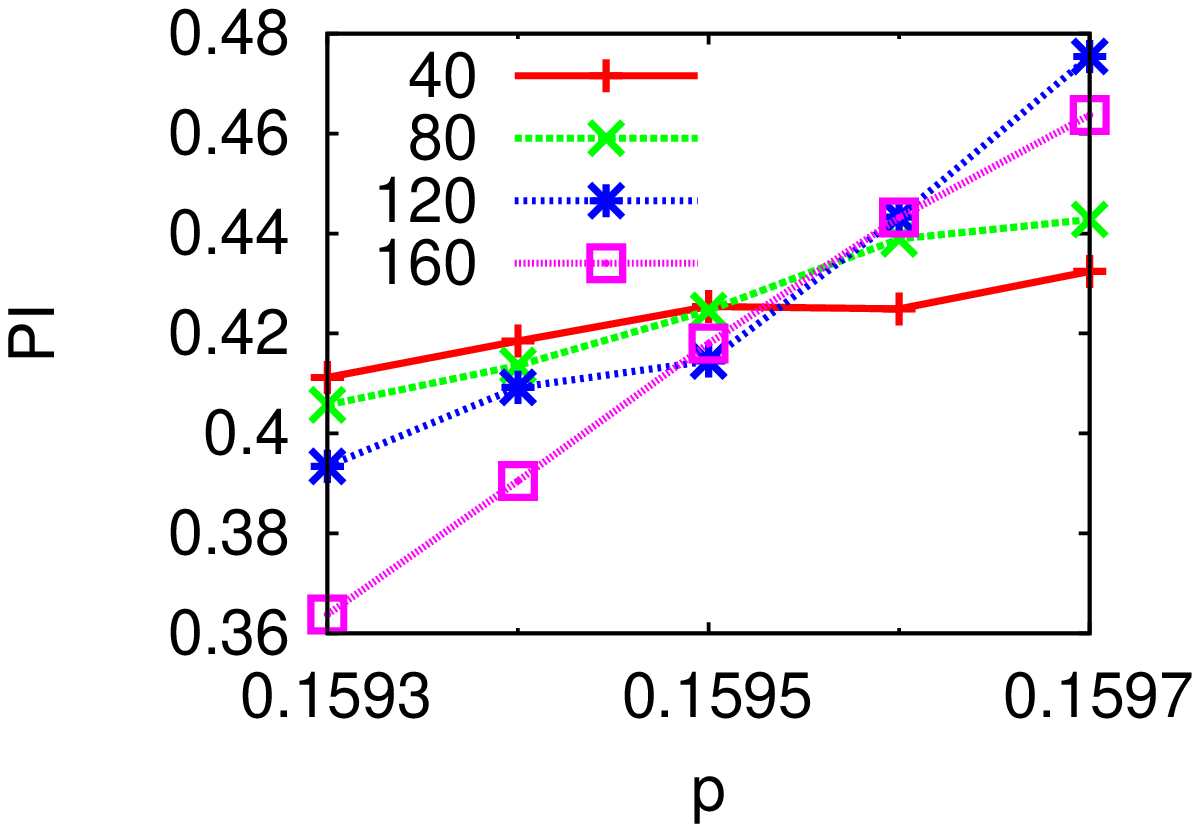}}
\subfigure{\includegraphics[width=0.23\textwidth]{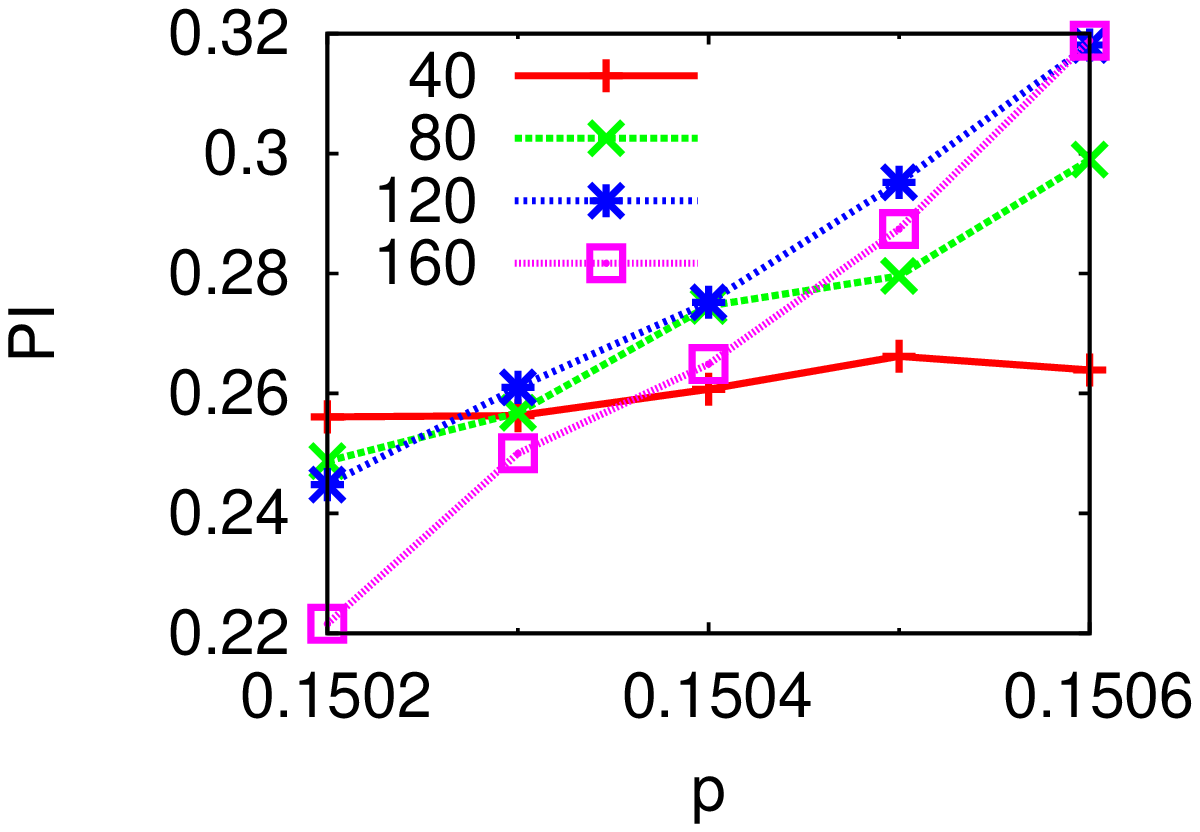}}
\subfigure{\includegraphics[width=0.23\textwidth]{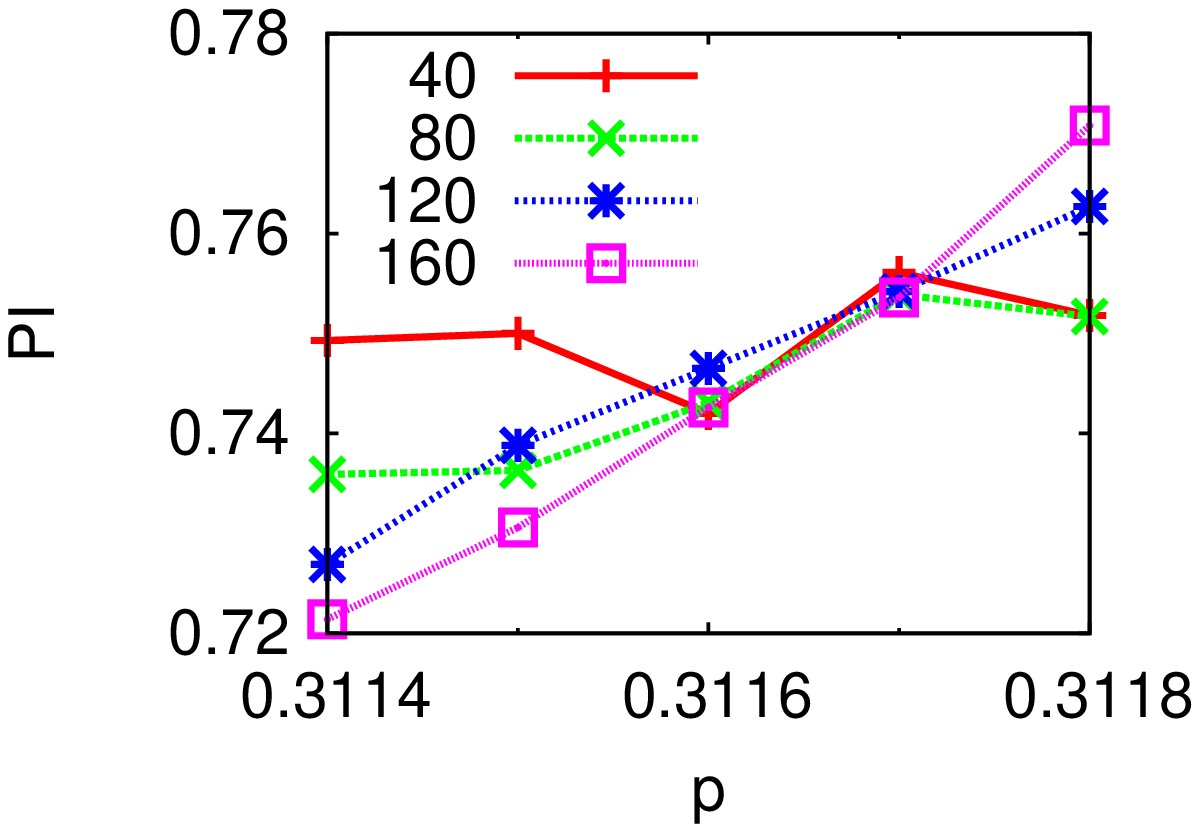}}
\addtocounter{subfigure}{-4}
\subfigure[4NN+3NN\label{fig-Lambda4N5N}]{\includegraphics[width=0.23\textwidth]{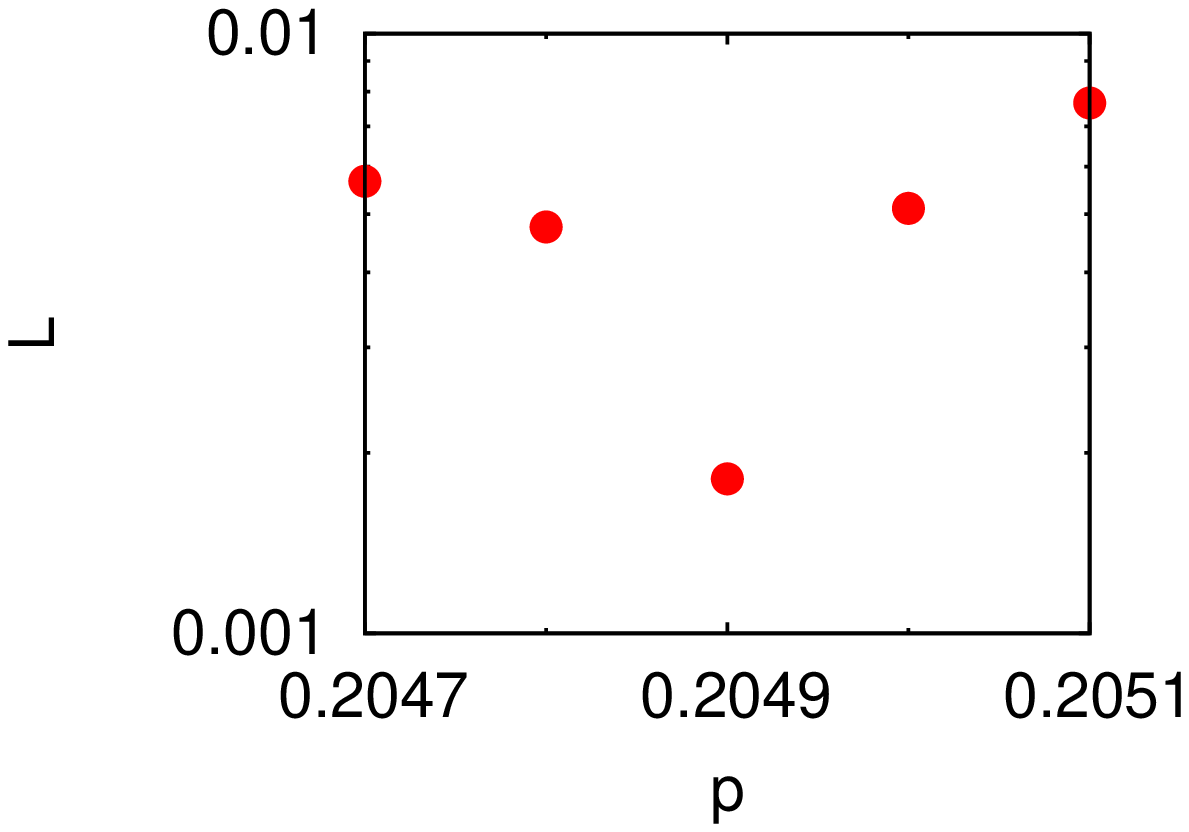}}
\subfigure[4NN+2NN\label{fig-Lambda3N5N}]{\includegraphics[width=0.23\textwidth]{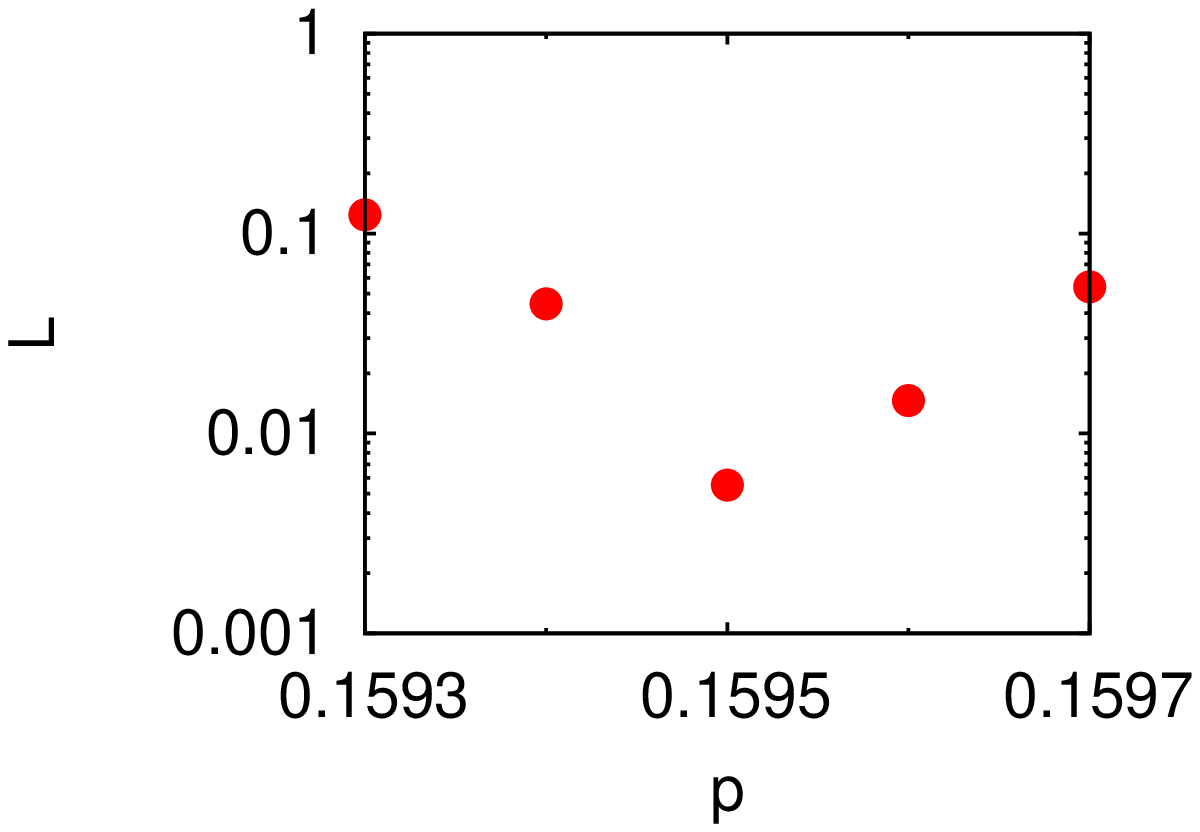}}
\subfigure[4NN+NN\label{fig-Lambda2N5N}]{\includegraphics[width=0.23\textwidth]{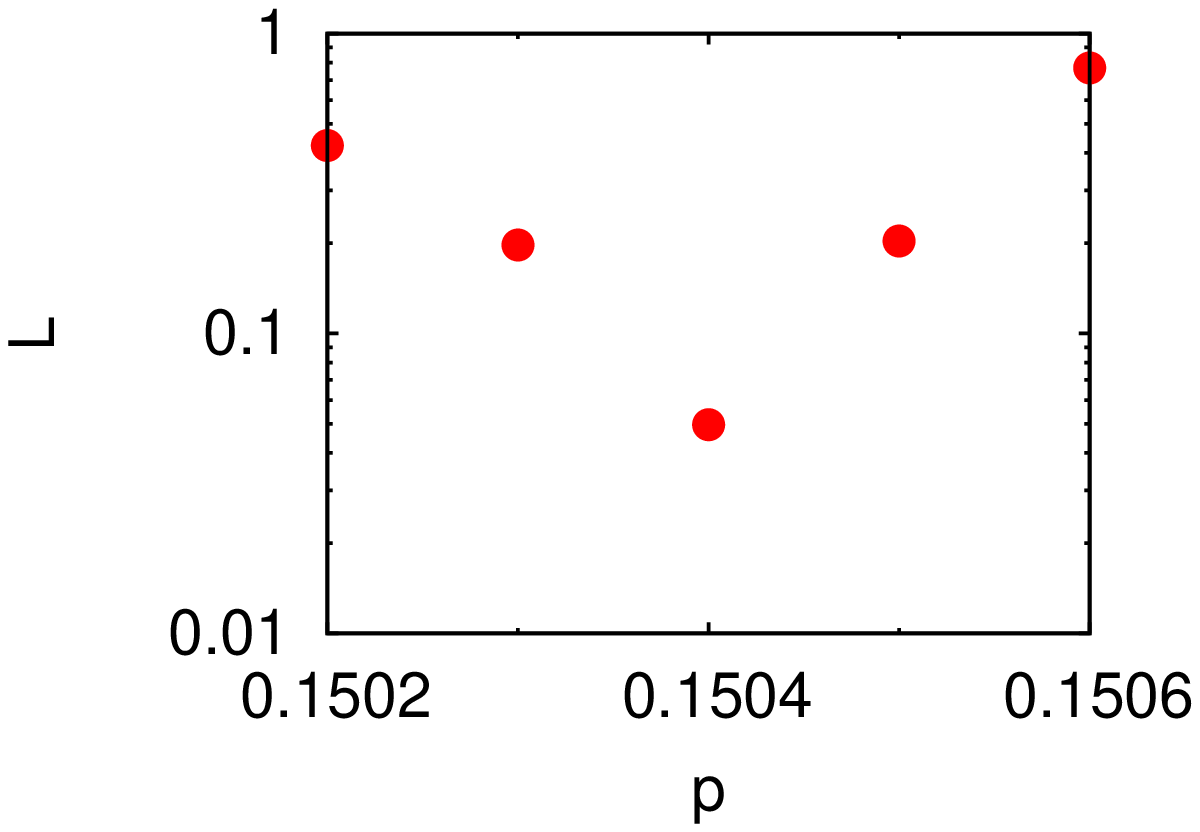}}
\subfigure[4NN\label{fig-Lambda5N}]{\includegraphics[width=0.23\textwidth]{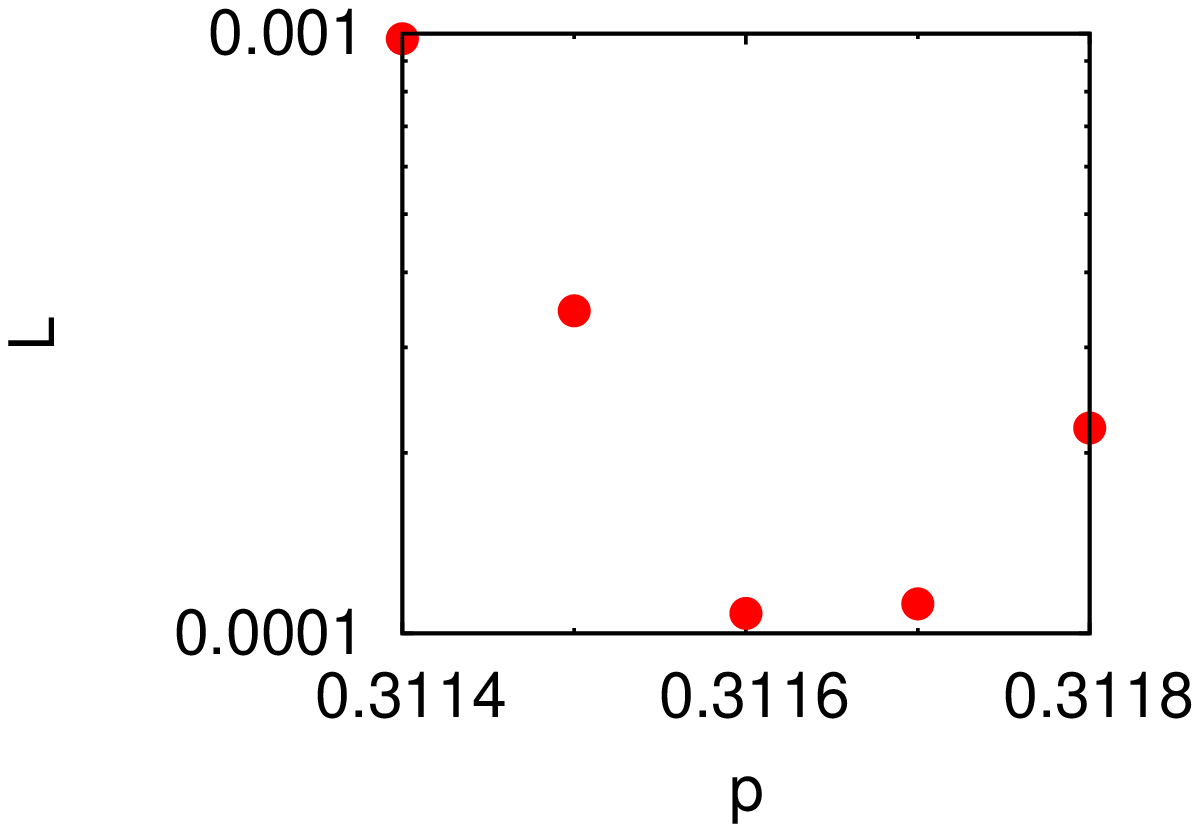}}
\caption{\label{fig-lambda} (Color online).
 Wrapping probability $W(p;L)$ and the pairwise sum $\lambda(p)$ vs. occupation probability $p$.
The results are averaged over $N_{\text{run}}=10^4$ runs.
The \change[Rev.2]{line colors (red, green, blue, violet)}{symbols ($+$, $\times$, $\xplus$, $\square$)} indicate the system linear sizes ($L=40$, 80, 120, 160), respectively.
The minima of $\lambda(p)$ correspond to the percolation thresholds $p_C$.}
\end{figure*}
%% ----------------------------------------------------------

For each pair $(p,L)$ of parameters $N_{\text{run}}=10^4$ lattices with randomly occupied $p\cdot L^3$ sites were simulated for $L=40$, 80, 120 and 160.
The wrapping probabilities $W(p;L)$ for various neighborhoods combinations are presented in Fig.~\ref{fig-W}.
\remove{These dependencies allow for the first approximation of the range $[p_1,p_2]$ locating the percolation threshold value $p_C$ with an accuracy
$\epsilon=|p_2-p_1|\approx 2\cdot 10^{-3}$. In Fig.~}
%%\ref{fig-W}
\remove{ the gray rectangles mark these  $[p_1,p_2]$ intervals.}

\remove{Now, with an accuracy higher by a factor of twenty we scan each of the intervals $[p_1,p_2]$ and plot the $\lambda(p)$ function near its minimum (see Fig.~}
%% \ref{fig-lambda}
\remove{).}
As it was mentioned in the Introduction, the results of computer simulations rather rarely reproduce a single common point of curves $W(p;L)$ unless the number $N_{\text{run}}$ of prepared lattices is very high.
It means, that finding the common point of $W(p;L)$ curves for various linear system sizes $L$ may be quite problematic.
In order to illustrate better this situation, we plot $W(p;L)$ dependencies near $p_C$ with sites occupation probability step $\Delta p=10^{-4}$ (see Fig.~\ref{fig-lambda}).
And indeed, except for 4NN+2NN+NN neighborhood, the curves $W(p;L)$ for various pairs of $L$ intersect at different points.
Moreover, for the smallest values of $L$ the dependencies $W(p;L)$ do not even increase monotonically with $p$.
At the same time, the dependencies $\lambda(p;L)$ prepared with the same accuracy $\Delta p=10^{-4}$ exhibit single and sharp minimum (it is worth to mention, that values of $\lambda$ are presented with the use of logarithmic scale).
The minimum of $\lambda(p)$ corresponds to the percolation threshold $p_C$. The estimated thresholds are presented in Tab.~\ref{tab-PT}.

The plots $W(p;L)$ presented in Fig.~\ref{fig-lambda}\add{ make possible to detrmine the length of interval where the true value of percolation threshold is located.
This length is equal to $\delta_W(p_C)=0.0004$.
Assuming that a real percolation threshold value is uniformly distributed in this interval one may evaluate the percolation threshold uncertainty as $u_W(p_C)=\delta_W(p_C)/\sqrt{3}\approx 0.00023$.
The approach based on $\lambda(p;L)$ dependence gives evaluation of $\delta_\lambda(p_C)$ twice smaller and consequently $u_\lambda(p_C)\approx 0.00012$.
On the other hand the method of $p_C$ estimation based solely on $W(p;L)$ dependencies, applied for similar neighborhood geometries leads to twice smaller lengths $\delta(p_C)$ and consequently similar uncertainties $u(p_C)$ but for ten times larger sampling} ($N_{\text{run}}=10^5$) \cite{Kurzawski2012}\add{.
One can conclude, that the method used by Bastas et al. leads to reaching uncertainty of the percolation threshold value $p_C$ similar to those obtained with classical method $u_\lambda(p_C)\approx u_W(p_C)$ but ten times faster.}

Note, that
an 
SC lattice with 4NN neighbors may be mapped
onto 
two independent interpenetrated SC lattices but with two
times larger lattice constant. Thus we expect
the
percolation threshold $p_C(\text{4NN})$ for the
next-next-next-next-nearest neighbors
to be equal
exactly
to $p_C(\text{NN})$. Indeed, the obtained value of
$p_C(4\text{NN})\approx${0.3116}\add[Rev.2]{0} agrees very well with values of
percolation threshold estimated for the nearest neighbors
$p_C(\text{NN})\approx 0.31160768(15)$ obtained very recently in
extensive Monte Carlo simulation \cite{Xu2014} and its earlier
estimations
\cite{PhysRevE.79.041141,*Lorenz1998,*Ballesteros1999,*PhysRevE.72.016126,*PhysRevE.87.052107,*Naeem1998,*Acharyya1998,*Grassberger1992}.

Note, however, that reaching such accuracy requires, for $L\le 128$,
sampling over $N_{\text{run}}=5\cdot 10^8$ lattices realization
\cite{Xu2014}, while we recovered
the first \change[Rev.2]{four}{five} digits of $p_C(\text{NN})$ with 
statistics lower by more than four orders of magnitude.

\add{Knowing percolation threshold may be practically useful for many systems with neighborhoods ranging beyond nearest neighbors}
\cite{Birenbaum,*Vandermeer,*Chen,*Bianconi,*Yiu,*Smits,*Wang,*Kriel,*Rodrigues}
\add{or next-nearest neighbors}
\cite{Gratens,*Albert,*Boer}.
\add{Thus practical application of $p_C$ values for longer ranges of interaction among systems' items cannot be generally excluded in all typical applications of the percolation theory, i.e. physics, chemistry, biology and social sciences.}

%% In literature known to us the percolation thresholds for SC random site percolation with sites outside Rubick's neighborhood are missing.
%% 0.311604(6),[84] = Grassberger1992
%% 0.311605(5),[85] = Acharyya1998
%% 0.311600(5),[86] = Naeem1998
%% 0.3116077(2),[83] = PhysRevE.87.052107
%% 0.3116077(4),[87] = PhysRevE.72.016126
%% 0.3116081(13),[88] = Ballesteros1999
%% 0.3116080(4),[89] = Lorenz1998
%% 0.3116004(35),[90] = PhysRevE.79.041141
%% ----------------------------------------------------------

\begin{acknowledgments}
The work was supported by the Polish Ministry of Science and Higher Education and its grants for scientific research and \href{http://projekt.plgrid.pl/en}{PL-Grid infrastructure}.
\end{acknowledgments}

\bibliography{percolation}

\begin{thebibliography}{71}%
\makeatletter
\providecommand \@ifxundefined [1]{%
 \@ifx{#1\undefined}
}%
\providecommand \@ifnum [1]{%
 \ifnum #1\expandafter \@firstoftwo
 \else \expandafter \@secondoftwo
 \fi
}%
\providecommand \@ifx [1]{%
 \ifx #1\expandafter \@firstoftwo
 \else \expandafter \@secondoftwo
 \fi
}%
\providecommand \natexlab [1]{#1}%
\providecommand \enquote  [1]{``#1''}%
\providecommand \bibnamefont  [1]{#1}%
\providecommand \bibfnamefont [1]{#1}%
\providecommand \citenamefont [1]{#1}%
\providecommand \href@noop [0]{\@secondoftwo}%
\providecommand \href [0]{\begingroup \@sanitize@url \@href}%
\providecommand \@href[1]{\@@startlink{#1}\@@href}%
\providecommand \@@href[1]{\endgroup#1\@@endlink}%
\providecommand \@sanitize@url [0]{\catcode `\\12\catcode `\$12\catcode
  `\&12\catcode `\#12\catcode `\^12\catcode `\_12\catcode `\%12\relax}%
\providecommand \@@startlink[1]{}%
\providecommand \@@endlink[0]{}%
\providecommand \url  [0]{\begingroup\@sanitize@url \@url }%
\providecommand \@url [1]{\endgroup\@href {#1}{\urlprefix }}%
\providecommand \urlprefix  [0]{URL }%
\providecommand \Eprint [0]{\href }%
\providecommand \doibase [0]{http://dx.doi.org/}%
\providecommand \selectlanguage [0]{\@gobble}%
\providecommand \bibinfo  [0]{\@secondoftwo}%
\providecommand \bibfield  [0]{\@secondoftwo}%
\providecommand \translation [1]{[#1]}%
\providecommand \BibitemOpen [0]{}%
\providecommand \bibitemStop [0]{}%
\providecommand \bibitemNoStop [0]{.\EOS\space}%
\providecommand \EOS [0]{\spacefactor3000\relax}%
\providecommand \BibitemShut  [1]{\csname bibitem#1\endcsname}%
\let\auto@bib@innerbib\@empty
%</preamble>
\bibitem [{\citenamefont {Stauffer}\ and\ \citenamefont
  {Aharony}(1994)}]{bookDS}%
  \BibitemOpen
  \bibfield  {author} {\bibinfo {author} {\bibfnamefont {D.}~\bibnamefont
  {Stauffer}}\ and\ \bibinfo {author} {\bibfnamefont {A.}~\bibnamefont
  {Aharony}},\ }\href@noop {} {\emph {\bibinfo {title} {Introduction to
  Percolation Theory}}},\ \bibinfo {edition} {2nd}\ ed.\ (\bibinfo  {publisher}
  {Taylor and Francis},\ \bibinfo {address} {London},\ \bibinfo {year}
  {1994})\BibitemShut {NoStop}%
\bibitem [{\citenamefont {Bollob\'as}\ and\ \citenamefont
  {Riordan}(2006)}]{bookBB}%
  \BibitemOpen
  \bibfield  {author} {\bibinfo {author} {\bibfnamefont {B.}~\bibnamefont
  {Bollob\'as}}\ and\ \bibinfo {author} {\bibfnamefont {O.}~\bibnamefont
  {Riordan}},\ }\href@noop {} {\emph {\bibinfo {title} {Percolation}}}\
  (\bibinfo  {publisher} {Cambridge UP},\ \bibinfo {address} {Cambridge},\
  \bibinfo {year} {2006})\BibitemShut {NoStop}%
\bibitem [{\citenamefont {Kesten}(1982)}]{bookHK}%
  \BibitemOpen
  \bibfield  {author} {\bibinfo {author} {\bibfnamefont {H.}~\bibnamefont
  {Kesten}},\ }\href@noop {} {\emph {\bibinfo {title} {Percolation Theory for
  Mathematicians}}}\ (\bibinfo  {publisher} {Brikhauser},\ \bibinfo {address}
  {Boston},\ \bibinfo {year} {1982})\BibitemShut {NoStop}%
\bibitem [{\citenamefont {Sahimi}(1994)}]{bookMS}%
  \BibitemOpen
  \bibfield  {author} {\bibinfo {author} {\bibfnamefont {M.}~\bibnamefont
  {Sahimi}},\ }\href@noop {} {\emph {\bibinfo {title} {Applications of
  Percolation Theory}}}\ (\bibinfo  {publisher} {Taylor and Francis},\ \bibinfo
  {address} {London},\ \bibinfo {year} {1994})\BibitemShut {NoStop}%
\bibitem [{\citenamefont {Broadbent}\ and\ \citenamefont
  {Hammersley}(1957)}]{Broadbent1957}%
  \BibitemOpen
  \bibfield  {author} {\bibinfo {author} {\bibfnamefont {S.~R.}\ \bibnamefont
  {Broadbent}}\ and\ \bibinfo {author} {\bibfnamefont {J.~M.}\ \bibnamefont
  {Hammersley}},\ }\href {\doibase 10.1017/S0305004100032680} {\bibfield
  {journal} {\bibinfo  {journal} {Mathematical Proceedings of the Cambridge
  Philosophical Society}\ }\textbf {\bibinfo {volume} {53}},\ \bibinfo {pages}
  {629} (\bibinfo {year} {1957})}\BibitemShut {NoStop}%
\bibitem [{\citenamefont {Frisch}\ \emph {et~al.}(1961)\citenamefont {Frisch},
  \citenamefont {Vyssotsky}, \citenamefont {Sonnenblick},\ and\ \citenamefont
  {Hammersley}}]{Frisch1961}%
  \BibitemOpen
  \bibfield  {author} {\bibinfo {author} {\bibfnamefont {H.}~\bibnamefont
  {Frisch}}, \bibinfo {author} {\bibfnamefont {V.}~\bibnamefont {Vyssotsky}},
  \bibinfo {author} {\bibfnamefont {E.}~\bibnamefont {Sonnenblick}}, \ and\
  \bibinfo {author} {\bibfnamefont {J.}~\bibnamefont {Hammersley}},\ }\href
  {\doibase 10.1103/PhysRev.124.1021} {\bibfield  {journal} {\bibinfo
  {journal} {Phys. Rev.}\ }\textbf {\bibinfo {volume} {124}},\ \bibinfo {pages}
  {1021} (\bibinfo {year} {1961})}\BibitemShut {NoStop}%
\bibitem [{\citenamefont {Frisch}\ \emph {et~al.}(1962)\citenamefont {Frisch},
  \citenamefont {Hammersley},\ and\ \citenamefont {Welsh}}]{Frisch1962}%
  \BibitemOpen
  \bibfield  {author} {\bibinfo {author} {\bibfnamefont {H.~L.}\ \bibnamefont
  {Frisch}}, \bibinfo {author} {\bibfnamefont {J.~M.}\ \bibnamefont
  {Hammersley}}, \ and\ \bibinfo {author} {\bibfnamefont {D.~J.~A.}\
  \bibnamefont {Welsh}},\ }\href {\doibase 10.1103/PhysRev.126.949} {\bibfield
  {journal} {\bibinfo  {journal} {Phys. Rev.}\ }\textbf {\bibinfo {volume}
  {126}},\ \bibinfo {pages} {949} (\bibinfo {year} {1962})}\BibitemShut
  {NoStop}%
\bibitem [{\citenamefont {Hoppe}\ and\ \citenamefont
  {Rodgers}(2014)}]{PhysRevE.90.012815}%
  \BibitemOpen
  \bibfield  {author} {\bibinfo {author} {\bibfnamefont {K.}~\bibnamefont
  {Hoppe}}\ and\ \bibinfo {author} {\bibfnamefont {G.~J.}\ \bibnamefont
  {Rodgers}},\ }\href {\doibase 10.1103/PhysRevE.90.012815} {\bibfield
  {journal} {\bibinfo  {journal} {Phys. Rev. E}\ }\textbf {\bibinfo {volume}
  {90}},\ \bibinfo {pages} {012815} (\bibinfo {year} {2014})}\BibitemShut
  {NoStop}%
\bibitem [{\citenamefont {Wierman}\ \emph {et~al.}(2005)\citenamefont
  {Wierman}, \citenamefont {Naor},\ and\ \citenamefont {Cheng}}]{Wierman2005}%
  \BibitemOpen
  \bibfield  {author} {\bibinfo {author} {\bibfnamefont {J.}~\bibnamefont
  {Wierman}}, \bibinfo {author} {\bibfnamefont {D.}~\bibnamefont {Naor}}, \
  and\ \bibinfo {author} {\bibfnamefont {R.}~\bibnamefont {Cheng}},\ }\href
  {\doibase 10.1103/PhysRevE.72.066116} {\bibfield  {journal} {\bibinfo
  {journal} {Phys. Rev. E}\ }\textbf {\bibinfo {volume} {72}},\ \bibinfo
  {pages} {066116} (\bibinfo {year} {2005})}\BibitemShut {NoStop}%
\bibitem [{\citenamefont {Rosowsky}(2000)}]{Rosowsky2000}%
  \BibitemOpen
  \bibfield  {author} {\bibinfo {author} {\bibfnamefont {A.}~\bibnamefont
  {Rosowsky}},\ }\href {\doibase 10.1007/s100510051101} {\bibfield  {journal}
  {\bibinfo  {journal} {Eur. Phys. J. B}\ }\textbf {\bibinfo {volume} {15}},\
  \bibinfo {pages} {77} (\bibinfo {year} {2000})}\BibitemShut {NoStop}%
\bibitem [{\citenamefont {Gaunt}\ and\ \citenamefont
  {Sykes}(1983)}]{ISI:A1983QF02000016}%
  \BibitemOpen
  \bibfield  {author} {\bibinfo {author} {\bibfnamefont {D.}~\bibnamefont
  {Gaunt}}\ and\ \bibinfo {author} {\bibfnamefont {M.}~\bibnamefont {Sykes}},\
  }\href {\doibase 10.1088/0305-4470/16/4/016} {\bibfield  {journal} {\bibinfo
  {journal} {J. Phys. A: Math. Gen.}\ }\textbf {\bibinfo {volume} {16}},\
  \bibinfo {pages} {783} (\bibinfo {year} {1983})}\BibitemShut {NoStop}%
\bibitem [{\citenamefont {Sykes}\ and\ \citenamefont
  {Glen}(1976)}]{ISI:A1976BD65000014}%
  \BibitemOpen
  \bibfield  {author} {\bibinfo {author} {\bibfnamefont {M.}~\bibnamefont
  {Sykes}}\ and\ \bibinfo {author} {\bibfnamefont {M.}~\bibnamefont {Glen}},\
  }\href {\doibase 10.1088/0305-4470/9/1/014} {\bibfield  {journal} {\bibinfo
  {journal} {J. Phys. A: Math. Gen.}\ }\textbf {\bibinfo {volume} {9}},\
  \bibinfo {pages} {87} (\bibinfo {year} {1976})}\BibitemShut {NoStop}%
\bibitem [{\citenamefont {Sykes}\ \emph
  {et~al.}(1976{\natexlab{a}})\citenamefont {Sykes}, \citenamefont {Gaunt},\
  and\ \citenamefont {Glen}}]{Sykes1976b}%
  \BibitemOpen
  \bibfield  {author} {\bibinfo {author} {\bibfnamefont {M.}~\bibnamefont
  {Sykes}}, \bibinfo {author} {\bibfnamefont {D.}~\bibnamefont {Gaunt}}, \ and\
  \bibinfo {author} {\bibfnamefont {M.}~\bibnamefont {Glen}},\ }\href {\doibase
  10.1088/0305-4470/9/1/015} {\bibfield  {journal} {\bibinfo  {journal} {J.
  Phys. A: Math. Gen.}\ }\textbf {\bibinfo {volume} {9}},\ \bibinfo {pages}
  {97} (\bibinfo {year} {1976}{\natexlab{a}})}\BibitemShut {NoStop}%
\bibitem [{\citenamefont {Sykes}\ \emph
  {et~al.}(1976{\natexlab{b}})\citenamefont {Sykes}, \citenamefont {Gaunt},\
  and\ \citenamefont {Glen}}]{Sykes1976c}%
  \BibitemOpen
  \bibfield  {author} {\bibinfo {author} {\bibfnamefont {M.}~\bibnamefont
  {Sykes}}, \bibinfo {author} {\bibfnamefont {D.}~\bibnamefont {Gaunt}}, \ and\
  \bibinfo {author} {\bibfnamefont {M.}~\bibnamefont {Glen}},\ }\href {\doibase
  10.1088/0305-4470/9/5/008} {\bibfield  {journal} {\bibinfo  {journal} {J.
  Phys. A: Math. Gen.}\ }\textbf {\bibinfo {volume} {9}},\ \bibinfo {pages}
  {715} (\bibinfo {year} {1976}{\natexlab{b}})}\BibitemShut {NoStop}%
\bibitem [{\citenamefont {Sykes}\ \emph
  {et~al.}(1976{\natexlab{c}})\citenamefont {Sykes}, \citenamefont {Gaunt},\
  and\ \citenamefont {Glen}}]{Sykes1976d}%
  \BibitemOpen
  \bibfield  {author} {\bibinfo {author} {\bibfnamefont {M.}~\bibnamefont
  {Sykes}}, \bibinfo {author} {\bibfnamefont {D.}~\bibnamefont {Gaunt}}, \ and\
  \bibinfo {author} {\bibfnamefont {M.}~\bibnamefont {Glen}},\ }\href {\doibase
  10.1088/0305-4470/9/5/009} {\bibfield  {journal} {\bibinfo  {journal} {J.
  Phys. A: Math. Gen.}\ }\textbf {\bibinfo {volume} {9}},\ \bibinfo {pages}
  {725} (\bibinfo {year} {1976}{\natexlab{c}})}\BibitemShut {NoStop}%
\bibitem [{\citenamefont {Gaunt}\ and\ \citenamefont
  {Sykes}(1976)}]{Gaunt1976}%
  \BibitemOpen
  \bibfield  {author} {\bibinfo {author} {\bibfnamefont {D.}~\bibnamefont
  {Gaunt}}\ and\ \bibinfo {author} {\bibfnamefont {M.}~\bibnamefont {Sykes}},\
  }\href {\doibase 10.1088/0305-4470/9/7/014} {\bibfield  {journal} {\bibinfo
  {journal} {J. Phys. A: Math. Gen.}\ }\textbf {\bibinfo {volume} {9}},\
  \bibinfo {pages} {1109} (\bibinfo {year} {1976})}\BibitemShut {NoStop}%
\bibitem [{\citenamefont {Clancy}\ \emph {et~al.}(2014)\citenamefont {Clancy},
  \citenamefont {Lupascu}, \citenamefont {Gretarsson}, \citenamefont {Islam},
  \citenamefont {Hu}, \citenamefont {Casa}, \citenamefont {Nelson},
  \citenamefont {LaMarra}, \citenamefont {Cao},\ and\ \citenamefont
  {Kim}}]{PhysRevB.89.054409}%
  \BibitemOpen
  \bibfield  {author} {\bibinfo {author} {\bibfnamefont {J.~P.}\ \bibnamefont
  {Clancy}}, \bibinfo {author} {\bibfnamefont {A.}~\bibnamefont {Lupascu}},
  \bibinfo {author} {\bibfnamefont {H.}~\bibnamefont {Gretarsson}}, \bibinfo
  {author} {\bibfnamefont {Z.}~\bibnamefont {Islam}}, \bibinfo {author}
  {\bibfnamefont {Y.~F.}\ \bibnamefont {Hu}}, \bibinfo {author} {\bibfnamefont
  {D.}~\bibnamefont {Casa}}, \bibinfo {author} {\bibfnamefont {C.~S.}\
  \bibnamefont {Nelson}}, \bibinfo {author} {\bibfnamefont {S.~C.}\
  \bibnamefont {LaMarra}}, \bibinfo {author} {\bibfnamefont {G.}~\bibnamefont
  {Cao}}, \ and\ \bibinfo {author} {\bibfnamefont {Y.-J.}\ \bibnamefont
  {Kim}},\ }\href {\doibase 10.1103/PhysRevB.89.054409} {\bibfield  {journal}
  {\bibinfo  {journal} {Phys. Rev. B}\ }\textbf {\bibinfo {volume} {89}},\
  \bibinfo {pages} {054409} (\bibinfo {year} {2014})}\BibitemShut {NoStop}%
\bibitem [{\citenamefont {Silva}\ \emph {et~al.}(2011)\citenamefont {Silva},
  \citenamefont {Simoes}, \citenamefont {Lanceros-Mendez},\ and\ \citenamefont
  {Vaia}}]{Silva2011}%
  \BibitemOpen
  \bibfield  {author} {\bibinfo {author} {\bibfnamefont {J.}~\bibnamefont
  {Silva}}, \bibinfo {author} {\bibfnamefont {R.}~\bibnamefont {Simoes}},
  \bibinfo {author} {\bibfnamefont {S.}~\bibnamefont {Lanceros-Mendez}}, \ and\
  \bibinfo {author} {\bibfnamefont {R.}~\bibnamefont {Vaia}},\ }\href {\doibase
  10.1209/0295-5075/93/37005} {\bibfield  {journal} {\bibinfo  {journal} {EPL}\
  }\textbf {\bibinfo {volume} {93}},\ \bibinfo {pages} {37005} (\bibinfo {year}
  {2011})}\BibitemShut {NoStop}%
\bibitem [{\citenamefont {Shearing}\ \emph {et~al.}(2010)\citenamefont
  {Shearing}, \citenamefont {Brett},\ and\ \citenamefont
  {Brandon}}]{Shearing2010}%
  \BibitemOpen
  \bibfield  {author} {\bibinfo {author} {\bibfnamefont {P.~R.}\ \bibnamefont
  {Shearing}}, \bibinfo {author} {\bibfnamefont {D.~J.~L.}\ \bibnamefont
  {Brett}}, \ and\ \bibinfo {author} {\bibfnamefont {N.~P.}\ \bibnamefont
  {Brandon}},\ }\href {\doibase 10.1179/095066010X12777205875679} {\bibfield
  {journal} {\bibinfo  {journal} {Int. Mater. Rev.}\ }\textbf {\bibinfo
  {volume} {55}},\ \bibinfo {pages} {347} (\bibinfo {year} {2010})}\BibitemShut
  {NoStop}%
\bibitem [{\citenamefont {Halperin}\ and\ \citenamefont
  {Bergman}(2010)}]{Halperin2010}%
  \BibitemOpen
  \bibfield  {author} {\bibinfo {author} {\bibfnamefont {B.~I.}\ \bibnamefont
  {Halperin}}\ and\ \bibinfo {author} {\bibfnamefont {D.~J.}\ \bibnamefont
  {Bergman}},\ }\href {\doibase 10.1016/j.physb.2010.01.002} {\bibfield
  {journal} {\bibinfo  {journal} {Physica B}\ }\textbf {\bibinfo {volume}
  {405}},\ \bibinfo {pages} {2908} (\bibinfo {year} {2010})}\BibitemShut
  {NoStop}%
\bibitem [{\citenamefont {Mun}\ \emph {et~al.}(2014)\citenamefont {Mun},
  \citenamefont {Kim}, \citenamefont {Prakashan}, \citenamefont {Jung},
  \citenamefont {Son},\ and\ \citenamefont {Park}}]{Mun2014}%
  \BibitemOpen
  \bibfield  {author} {\bibinfo {author} {\bibfnamefont {S.~C.}\ \bibnamefont
  {Mun}}, \bibinfo {author} {\bibfnamefont {M.}~\bibnamefont {Kim}}, \bibinfo
  {author} {\bibfnamefont {K.}~\bibnamefont {Prakashan}}, \bibinfo {author}
  {\bibfnamefont {H.~J.}\ \bibnamefont {Jung}}, \bibinfo {author}
  {\bibfnamefont {Y.}~\bibnamefont {Son}}, \ and\ \bibinfo {author}
  {\bibfnamefont {O.~O.}\ \bibnamefont {Park}},\ }\href {\doibase
  10.1016/j.carbon.2013.09.056} {\bibfield  {journal} {\bibinfo  {journal}
  {Carbon}\ }\textbf {\bibinfo {volume} {67}},\ \bibinfo {pages} {64} (\bibinfo
  {year} {2014})}\BibitemShut {NoStop}%
\bibitem [{\citenamefont {Amiaz}\ \emph {et~al.}(2011)\citenamefont {Amiaz},
  \citenamefont {Sorek}, \citenamefont {Enzel},\ and\ \citenamefont
  {Dahan}}]{Amiaz2011}%
  \BibitemOpen
  \bibfield  {author} {\bibinfo {author} {\bibfnamefont {Y.}~\bibnamefont
  {Amiaz}}, \bibinfo {author} {\bibfnamefont {S.}~\bibnamefont {Sorek}},
  \bibinfo {author} {\bibfnamefont {Y.}~\bibnamefont {Enzel}}, \ and\ \bibinfo
  {author} {\bibfnamefont {O.}~\bibnamefont {Dahan}},\ }\href {\doibase
  10.1029/2011WR010747} {\bibfield  {journal} {\bibinfo  {journal} {Water
  Resour. Res.}\ }\textbf {\bibinfo {volume} {47}},\ \bibinfo {pages} {W10513}
  (\bibinfo {year} {2011})}\BibitemShut {NoStop}%
\bibitem [{\citenamefont {Bolandtaba}\ and\ \citenamefont
  {Skauge}(2011)}]{Bolandtaba2011}%
  \BibitemOpen
  \bibfield  {author} {\bibinfo {author} {\bibfnamefont {S.~F.}\ \bibnamefont
  {Bolandtaba}}\ and\ \bibinfo {author} {\bibfnamefont {A.}~\bibnamefont
  {Skauge}},\ }\href {\doibase 10.1007/s11242-011-9775-0} {\bibfield  {journal}
  {\bibinfo  {journal} {Transp. Porous Media}\ }\textbf {\bibinfo {volume}
  {89}},\ \bibinfo {pages} {357} (\bibinfo {year} {2011})}\BibitemShut
  {NoStop}%
\bibitem [{\citenamefont {Mourzenko}\ \emph {et~al.}(2011)\citenamefont
  {Mourzenko}, \citenamefont {Thovert},\ and\ \citenamefont
  {Adler}}]{Mourzenko2011}%
  \BibitemOpen
  \bibfield  {author} {\bibinfo {author} {\bibfnamefont {V.~V.}\ \bibnamefont
  {Mourzenko}}, \bibinfo {author} {\bibfnamefont {J.~F.}\ \bibnamefont
  {Thovert}}, \ and\ \bibinfo {author} {\bibfnamefont {P.~M.}\ \bibnamefont
  {Adler}},\ }\href {\doibase 10.1103/PhysRevE.84.036307} {\bibfield  {journal}
  {\bibinfo  {journal} {Phys. Rev. E}\ }\textbf {\bibinfo {volume} {84}},\
  \bibinfo {pages} {036307} (\bibinfo {year} {2011})}\BibitemShut {NoStop}%
\bibitem [{\citenamefont {Abades}\ \emph {et~al.}(2014)\citenamefont {Abades},
  \citenamefont {Gaxiola},\ and\ \citenamefont {Marquet}}]{Abades2014}%
  \BibitemOpen
  \bibfield  {author} {\bibinfo {author} {\bibfnamefont {S.~R.}\ \bibnamefont
  {Abades}}, \bibinfo {author} {\bibfnamefont {A.}~\bibnamefont {Gaxiola}}, \
  and\ \bibinfo {author} {\bibfnamefont {P.~A.}\ \bibnamefont {Marquet}},\
  }\href {\doibase 10.1111/1365-2745.12321} {\bibfield  {journal} {\bibinfo
  {journal} {J. Ecology}\ }\textbf {\bibinfo {volume} {102}},\ \bibinfo {pages}
  {1386} (\bibinfo {year} {2014})}\BibitemShut {NoStop}%
\bibitem [{\citenamefont {Camelo-Neto}\ and\ \citenamefont
  {Coutinho}(2011)}]{Camelo-Neto2011}%
  \BibitemOpen
  \bibfield  {author} {\bibinfo {author} {\bibfnamefont {G.}~\bibnamefont
  {Camelo-Neto}}\ and\ \bibinfo {author} {\bibfnamefont {S.}~\bibnamefont
  {Coutinho}},\ }\href {\doibase 10.1088/1742-5468/2011/06/P06018} {\bibfield
  {journal} {\bibinfo  {journal} {J. Stat. Mech.-Theory Exp.}\ }\textbf
  {\bibinfo {volume} {2011}},\ \bibinfo {pages} {P06018} (\bibinfo {year}
  {2011})}\BibitemShut {NoStop}%
\bibitem [{\citenamefont {Guisoni}\ \emph {et~al.}(2011)\citenamefont
  {Guisoni}, \citenamefont {Loscar},\ and\ \citenamefont
  {Albano}}]{Guisoni2011}%
  \BibitemOpen
  \bibfield  {author} {\bibinfo {author} {\bibfnamefont {N.}~\bibnamefont
  {Guisoni}}, \bibinfo {author} {\bibfnamefont {E.~S.}\ \bibnamefont {Loscar}},
  \ and\ \bibinfo {author} {\bibfnamefont {E.~V.}\ \bibnamefont {Albano}},\
  }\href {\doibase 10.1103/PhysRevE.83.011125} {\bibfield  {journal} {\bibinfo
  {journal} {Phys. Rev. E}\ }\textbf {\bibinfo {volume} {83}},\ \bibinfo
  {pages} {011125} (\bibinfo {year} {2011})}\BibitemShut {NoStop}%
\bibitem [{\citenamefont {Simeoni}\ \emph {et~al.}(2011)\citenamefont
  {Simeoni}, \citenamefont {Salinesi},\ and\ \citenamefont
  {Morandini}}]{Simeoni2011}%
  \BibitemOpen
  \bibfield  {author} {\bibinfo {author} {\bibfnamefont {A.}~\bibnamefont
  {Simeoni}}, \bibinfo {author} {\bibfnamefont {P.}~\bibnamefont {Salinesi}}, \
  and\ \bibinfo {author} {\bibfnamefont {F.}~\bibnamefont {Morandini}},\ }\href
  {\doibase 10.1071/WF09006} {\bibfield  {journal} {\bibinfo  {journal} {Int.
  J. Wildland Fire}\ }\textbf {\bibinfo {volume} {20}},\ \bibinfo {pages} {625}
  (\bibinfo {year} {2011})}\BibitemShut {NoStop}%
\bibitem [{\citenamefont {Malarz}\ \emph {et~al.}(2002)\citenamefont {Malarz},
  \citenamefont {Kaczanowska},\ and\ \citenamefont
  {Kulakowski}}]{Kaczanowska2002}%
  \BibitemOpen
  \bibfield  {author} {\bibinfo {author} {\bibfnamefont {K.}~\bibnamefont
  {Malarz}}, \bibinfo {author} {\bibfnamefont {S.}~\bibnamefont {Kaczanowska}},
  \ and\ \bibinfo {author} {\bibfnamefont {K.}~\bibnamefont {Kulakowski}},\
  }\href {\doibase 10.1142/S0129183102003760} {\bibfield  {journal} {\bibinfo
  {journal} {Int. J. Mod. Phys. C}\ }\textbf {\bibinfo {volume} {13}},\
  \bibinfo {pages} {1017} (\bibinfo {year} {2002})}\BibitemShut {NoStop}%
\bibitem [{\citenamefont {Silverberg}\ \emph {et~al.}(2014)\citenamefont
  {Silverberg}, \citenamefont {Barrett}, \citenamefont {Das}, \citenamefont
  {Petersen}, \citenamefont {Bonassar},\ and\ \citenamefont
  {Cohen}}]{Silverberg2014}%
  \BibitemOpen
  \bibfield  {author} {\bibinfo {author} {\bibfnamefont {J.}~\bibnamefont
  {Silverberg}}, \bibinfo {author} {\bibfnamefont {A.}~\bibnamefont {Barrett}},
  \bibinfo {author} {\bibfnamefont {M.}~\bibnamefont {Das}}, \bibinfo {author}
  {\bibfnamefont {P.}~\bibnamefont {Petersen}}, \bibinfo {author}
  {\bibfnamefont {L.}~\bibnamefont {Bonassar}}, \ and\ \bibinfo {author}
  {\bibfnamefont {I.}~\bibnamefont {Cohen}},\ }\href {\doibase
  10.1016/j.bpj.2014.08.011} {\bibfield  {journal} {\bibinfo  {journal}
  {Biophys. J.}\ }\textbf {\bibinfo {volume} {107}},\ \bibinfo {pages} {1721}
  (\bibinfo {year} {2014})}\BibitemShut {NoStop}%
\bibitem [{\citenamefont {Suzuki}\ and\ \citenamefont
  {Sasaki}(2011)}]{Suzuki2011}%
  \BibitemOpen
  \bibfield  {author} {\bibinfo {author} {\bibfnamefont {S.~U.}\ \bibnamefont
  {Suzuki}}\ and\ \bibinfo {author} {\bibfnamefont {A.}~\bibnamefont
  {Sasaki}},\ }\href {\doibase 10.1016/j.jtbi.2011.02.002} {\bibfield
  {journal} {\bibinfo  {journal} {J. Theor. Biol.}\ }\textbf {\bibinfo {volume}
  {276}},\ \bibinfo {pages} {117} (\bibinfo {year} {2011})}\BibitemShut
  {NoStop}%
\bibitem [{\citenamefont {Lindquist}\ \emph {et~al.}(2011)\citenamefont
  {Lindquist}, \citenamefont {Ma}, \citenamefont {van~den Driessche},\ and\
  \citenamefont {Willeboordse}}]{Lindquist2011}%
  \BibitemOpen
  \bibfield  {author} {\bibinfo {author} {\bibfnamefont {J.}~\bibnamefont
  {Lindquist}}, \bibinfo {author} {\bibfnamefont {J.}~\bibnamefont {Ma}},
  \bibinfo {author} {\bibfnamefont {P.}~\bibnamefont {van~den Driessche}}, \
  and\ \bibinfo {author} {\bibfnamefont {F.~H.}\ \bibnamefont {Willeboordse}},\
  }\href {\doibase 10.1007/s00285-010-0331-2} {\bibfield  {journal} {\bibinfo
  {journal} {J. Math. Biol.}\ }\textbf {\bibinfo {volume} {62}},\ \bibinfo
  {pages} {143} (\bibinfo {year} {2011})}\BibitemShut {NoStop}%
\bibitem [{\citenamefont {Naumova}\ \emph {et~al.}(2008)\citenamefont
  {Naumova}, \citenamefont {Gorski},\ and\ \citenamefont
  {Naumov}}]{Naumova2008}%
  \BibitemOpen
  \bibfield  {author} {\bibinfo {author} {\bibfnamefont {E.~N.}\ \bibnamefont
  {Naumova}}, \bibinfo {author} {\bibfnamefont {J.}~\bibnamefont {Gorski}}, \
  and\ \bibinfo {author} {\bibfnamefont {Y.~N.}\ \bibnamefont {Naumov}},\
  }\href@noop {} {\bibfield  {journal} {\bibinfo  {journal} {Ann. Zool. Fenn.}\
  }\textbf {\bibinfo {volume} {45}},\ \bibinfo {pages} {369} (\bibinfo {year}
  {2008})}\BibitemShut {NoStop}%
\bibitem [{\citenamefont {Floyd}\ \emph {et~al.}(2008)\citenamefont {Floyd},
  \citenamefont {Kay},\ and\ \citenamefont {Shapiro}}]{Floyd2008}%
  \BibitemOpen
  \bibfield  {author} {\bibinfo {author} {\bibfnamefont {W.}~\bibnamefont
  {Floyd}}, \bibinfo {author} {\bibfnamefont {L.}~\bibnamefont {Kay}}, \ and\
  \bibinfo {author} {\bibfnamefont {M.}~\bibnamefont {Shapiro}},\ }\href
  {\doibase 10.1007/s11538-007-9275-0} {\bibfield  {journal} {\bibinfo
  {journal} {Bull. Math. Biol.}\ }\textbf {\bibinfo {volume} {70}},\ \bibinfo
  {pages} {713} (\bibinfo {year} {2008})}\BibitemShut {NoStop}%
\bibitem [{\citenamefont {Chandrashekar}(2014)}]{Chandrashekar2014}%
  \BibitemOpen
  \bibfield  {author} {\bibinfo {author} {\bibfnamefont {T.}~\bibnamefont
  {Chandrashekar}, \bibfnamefont {C.~M.~Busch}},\ }\href {\doibase
  10.1038/srep06583} {\bibfield  {journal} {\bibinfo  {journal} {Sci. Rep.}\
  }\textbf {\bibinfo {volume} {4}},\ \bibinfo {pages} {6583} (\bibinfo {year}
  {2014})}\BibitemShut {NoStop}%
\bibitem [{\citenamefont {Fisher}(1971)}]{Fisher1971}%
  \BibitemOpen
  \bibfield  {author} {\bibinfo {author} {\bibfnamefont {M.~E.}\ \bibnamefont
  {Fisher}},\ }in\ \href@noop {} {\emph {\bibinfo {booktitle} {Critical
  Phenomena}}},\ \bibinfo {editor} {edited by\ \bibinfo {editor} {\bibfnamefont
  {M.~S.}\ \bibnamefont {Green}}}\ (\bibinfo  {publisher} {Academic Press},\
  \bibinfo {address} {London},\ \bibinfo {year} {1971})\ p.~\bibinfo {pages}
  {1}\BibitemShut {NoStop}%
\bibitem [{\citenamefont {Privman}(1990)}]{bookVP}%
  \BibitemOpen
  \bibfield  {author} {\bibinfo {author} {\bibfnamefont {V.}~\bibnamefont
  {Privman}},\ }in\ \href@noop {} {\emph {\bibinfo {booktitle} {Finite size
  scaling and numerical simulation of statistical systems}}},\ \bibinfo
  {editor} {edited by\ \bibinfo {editor} {\bibfnamefont {V.}~\bibnamefont
  {Privman}}}\ (\bibinfo  {publisher} {World Scientific},\ \bibinfo {address}
  {Singapore},\ \bibinfo {year} {1990})\ p.~\bibinfo {pages} {1}\BibitemShut
  {NoStop}%
\bibitem [{\citenamefont {Binder}(1992)}]{Binder1992}%
  \BibitemOpen
  \bibfield  {author} {\bibinfo {author} {\bibfnamefont {K.}~\bibnamefont
  {Binder}},\ }in\ \href@noop {} {\emph {\bibinfo {booktitle} {Computational
  Methods in Field Theory}}},\ \bibinfo {editor} {edited by\ \bibinfo {editor}
  {\bibfnamefont {C.~B.}\ \bibnamefont {Lang}}\ and\ \bibinfo {editor}
  {\bibfnamefont {H.}~\bibnamefont {Gausterer}}}\ (\bibinfo  {publisher}
  {Springer},\ \bibinfo {address} {Berlin},\ \bibinfo {year}
  {1992})\BibitemShut {NoStop}%
\bibitem [{\citenamefont {Landau}\ and\ \citenamefont {Binder}(2005)}]{bookDL}%
  \BibitemOpen
  \bibfield  {author} {\bibinfo {author} {\bibfnamefont {D.~P.}\ \bibnamefont
  {Landau}}\ and\ \bibinfo {author} {\bibfnamefont {K.}~\bibnamefont
  {Binder}},\ }\href@noop {} {\emph {\bibinfo {title} {A Guide to Monte Carlo
  Simulations in Statistical Physics}}},\ \bibinfo {edition} {2nd}\ ed.\
  (\bibinfo  {publisher} {Cambridge UP},\ \bibinfo {address} {Cambridge},\
  \bibinfo {year} {2005})\BibitemShut {NoStop}%
\bibitem [{\citenamefont {Bastas}\ \emph {et~al.}(2011)\citenamefont {Bastas},
  \citenamefont {Kosmidis},\ and\ \citenamefont
  {Argyrakis}}]{PhysRevE.84.066112}%
  \BibitemOpen
  \bibfield  {author} {\bibinfo {author} {\bibfnamefont {N.}~\bibnamefont
  {Bastas}}, \bibinfo {author} {\bibfnamefont {K.}~\bibnamefont {Kosmidis}}, \
  and\ \bibinfo {author} {\bibfnamefont {P.}~\bibnamefont {Argyrakis}},\ }\href
  {\doibase 10.1103/PhysRevE.84.066112} {\bibfield  {journal} {\bibinfo
  {journal} {Phys. Rev. E}\ }\textbf {\bibinfo {volume} {84}},\ \bibinfo
  {pages} {066112} (\bibinfo {year} {2011})}\BibitemShut {NoStop}%
\bibitem [{\citenamefont {Bastas}\ \emph {et~al.}(2014)\citenamefont {Bastas},
  \citenamefont {Kosmidis}, \citenamefont {Giazitzidis},\ and\ \citenamefont
  {Maragakis}}]{Bastas2014}%
  \BibitemOpen
  \bibfield  {author} {\bibinfo {author} {\bibfnamefont {N.}~\bibnamefont
  {Bastas}}, \bibinfo {author} {\bibfnamefont {K.}~\bibnamefont {Kosmidis}},
  \bibinfo {author} {\bibfnamefont {P.}~\bibnamefont {Giazitzidis}}, \ and\
  \bibinfo {author} {\bibfnamefont {M.}~\bibnamefont {Maragakis}},\ }\href@noop
  {} {\enquote {\bibinfo {title} {Method for estimating critical exponents in
  percolation processes with low sampling},}\ } (\bibinfo {year} {2014}),\
  \Eprint {http://arxiv.org/abs/1411.5834} {arXiv:1411.5834
  [cond-mat.stat-mech]} \BibitemShut {NoStop}%
\bibitem [{\citenamefont {Malarz}\ and\ \citenamefont
  {Galam}(2005)}]{Galam2005a}%
  \BibitemOpen
  \bibfield  {author} {\bibinfo {author} {\bibfnamefont {K.}~\bibnamefont
  {Malarz}}\ and\ \bibinfo {author} {\bibfnamefont {S.}~\bibnamefont {Galam}},\
  }\href {\doibase 10.1103/PhysRevE.71.016125} {\bibfield  {journal} {\bibinfo
  {journal} {Phys. Rev. E}\ }\textbf {\bibinfo {volume} {71}},\ \bibinfo
  {pages} {016125} (\bibinfo {year} {2005})}\BibitemShut {NoStop}%
\bibitem [{\citenamefont {Galam}\ and\ \citenamefont
  {Malarz}(2005)}]{Galam2005b}%
  \BibitemOpen
  \bibfield  {author} {\bibinfo {author} {\bibfnamefont {S.}~\bibnamefont
  {Galam}}\ and\ \bibinfo {author} {\bibfnamefont {K.}~\bibnamefont {Malarz}},\
  }\href {\doibase 10.1103/PhysRevE.72.027103} {\bibfield  {journal} {\bibinfo
  {journal} {Phys. Rev. E}\ }\textbf {\bibinfo {volume} {72}},\ \bibinfo
  {pages} {027103} (\bibinfo {year} {2005})}\BibitemShut {NoStop}%
\bibitem [{\citenamefont {Majewski}\ and\ \citenamefont
  {Malarz}(2007)}]{Majewski2007}%
  \BibitemOpen
  \bibfield  {author} {\bibinfo {author} {\bibfnamefont {M.}~\bibnamefont
  {Majewski}}\ and\ \bibinfo {author} {\bibfnamefont {K.}~\bibnamefont
  {Malarz}},\ }\href
  {http://www.actaphys.uj.edu.pl/_cur/store/vol38/pdf/v38p2191.pdf} {\bibfield
  {journal} {\bibinfo  {journal} {Acta Phys. Pol. B}\ }\textbf {\bibinfo
  {volume} {38}},\ \bibinfo {pages} {2191} (\bibinfo {year}
  {2007})}\BibitemShut {NoStop}%
\bibitem [{\citenamefont {Kurzawski}\ and\ \citenamefont
  {Malarz}(2012)}]{Kurzawski2012}%
  \BibitemOpen
  \bibfield  {author} {\bibinfo {author} {\bibfnamefont {{\L}.}~\bibnamefont
  {Kurzawski}}\ and\ \bibinfo {author} {\bibfnamefont {K.}~\bibnamefont
  {Malarz}},\ }\href {\doibase 10.1016/S0034-4877(12)60036-6} {\bibfield
  {journal} {\bibinfo  {journal} {Rep. Math. Phys.}\ }\textbf {\bibinfo
  {volume} {70}},\ \bibinfo {pages} {163} (\bibinfo {year} {2012})}\BibitemShut
  {NoStop}%
\bibitem [{Note1()}]{Note1}%
  \BibitemOpen
  \bibinfo {note} {The detailed studies of properties $W(p;L)$ and its scaling
  function $\protect \mathcal {G}(\cdot )$ but for a square lattice are given
  in Ref.~\cite {Newman2001}.}\BibitemShut {Stop}%
\bibitem [{\citenamefont {Newman}\ and\ \citenamefont
  {Ziff}(2001)}]{Newman2001}%
  \BibitemOpen
  \bibfield  {author} {\bibinfo {author} {\bibfnamefont {M.}~\bibnamefont
  {Newman}}\ and\ \bibinfo {author} {\bibfnamefont {R.}~\bibnamefont {Ziff}},\
  }\href {\doibase 10.1103/PhysRevE.64.016706} {\bibfield  {journal} {\bibinfo
  {journal} {Phys. Rev. E}\ }\textbf {\bibinfo {volume} {64}},\ \bibinfo
  {pages} {016706} (\bibinfo {year} {2001})}\BibitemShut {NoStop}%
\bibitem [{\citenamefont {Hoshen}\ and\ \citenamefont
  {Kopelman}(1976)}]{Hoshen1976a}%
  \BibitemOpen
  \bibfield  {author} {\bibinfo {author} {\bibfnamefont {J.}~\bibnamefont
  {Hoshen}}\ and\ \bibinfo {author} {\bibfnamefont {R.}~\bibnamefont
  {Kopelman}},\ }\href {\doibase 10.1103/PhysRevB.14.3438} {\bibfield
  {journal} {\bibinfo  {journal} {Phys. Rev. B}\ }\textbf {\bibinfo {volume}
  {14}},\ \bibinfo {pages} {3438} (\bibinfo {year} {1976})}\BibitemShut
  {NoStop}%
\bibitem [{\citenamefont {Leath}(1976)}]{Leath1976}%
  \BibitemOpen
  \bibfield  {author} {\bibinfo {author} {\bibfnamefont {P.~L.}\ \bibnamefont
  {Leath}},\ }\href {\doibase 10.1103/PhysRevLett.36.921} {\bibfield  {journal}
  {\bibinfo  {journal} {Phys. Rev. Lett.}\ }\textbf {\bibinfo {volume} {36}},\
  \bibinfo {pages} {921} (\bibinfo {year} {1976})}\BibitemShut {NoStop}%
\bibitem [{\citenamefont {Torin}(2014)}]{Torin2014}%
  \BibitemOpen
  \bibfield  {author} {\bibinfo {author} {\bibfnamefont {I.~V.}\ \bibnamefont
  {Torin}},\ }\href {\doibase 10.1142/S0129183114500648} {\bibfield  {journal}
  {\bibinfo  {journal} {Int. J. Mod. Phys. C}\ }\textbf {\bibinfo {volume}
  {25}},\ \bibinfo {pages} {1450064} (\bibinfo {year} {2014})}\BibitemShut
  {NoStop}%
\bibitem [{\citenamefont {Xu}\ \emph {et~al.}(2014)\citenamefont {Xu},
  \citenamefont {Wang}, \citenamefont {Lv},\ and\ \citenamefont
  {Deng}}]{Xu2014}%
  \BibitemOpen
  \bibfield  {author} {\bibinfo {author} {\bibfnamefont {X.}~\bibnamefont
  {Xu}}, \bibinfo {author} {\bibfnamefont {J.}~\bibnamefont {Wang}}, \bibinfo
  {author} {\bibfnamefont {J.-P.}\ \bibnamefont {Lv}}, \ and\ \bibinfo {author}
  {\bibfnamefont {Y.}~\bibnamefont {Deng}},\ }\href {\doibase
  10.1007/s11467-013-0403-z} {\bibfield  {journal} {\bibinfo  {journal} {Front.
  Phys.}\ }\textbf {\bibinfo {volume} {9}},\ \bibinfo {pages} {113} (\bibinfo
  {year} {2014})}\BibitemShut {NoStop}%
\bibitem [{\citenamefont {\v{S}kvor}\ and\ \citenamefont
  {Nezbeda}(2009)}]{PhysRevE.79.041141}%
  \BibitemOpen
  \bibfield  {author} {\bibinfo {author} {\bibfnamefont {J.}~\bibnamefont
  {\v{S}kvor}}\ and\ \bibinfo {author} {\bibfnamefont {I.}~\bibnamefont
  {Nezbeda}},\ }\href {\doibase 10.1103/PhysRevE.79.041141} {\bibfield
  {journal} {\bibinfo  {journal} {Phys. Rev. E}\ }\textbf {\bibinfo {volume}
  {79}},\ \bibinfo {pages} {041141} (\bibinfo {year} {2009})}\BibitemShut
  {NoStop}%
\bibitem [{\citenamefont {Lorenz}\ and\ \citenamefont
  {Ziff}(1998)}]{Lorenz1998}%
  \BibitemOpen
  \bibfield  {author} {\bibinfo {author} {\bibfnamefont {C.~D.}\ \bibnamefont
  {Lorenz}}\ and\ \bibinfo {author} {\bibfnamefont {R.~M.}\ \bibnamefont
  {Ziff}},\ }\href {http://stacks.iop.org/0305-4470/31/i=40/a=009} {\bibfield
  {journal} {\bibinfo  {journal} {J. Phys. A: Math. Gen.}\ }\textbf {\bibinfo
  {volume} {31}},\ \bibinfo {pages} {8147} (\bibinfo {year}
  {1998})}\BibitemShut {NoStop}%
\bibitem [{\citenamefont {Ballesteros}\ \emph {et~al.}(1999)\citenamefont
  {Ballesteros}, \citenamefont {Fern\'andez}, \citenamefont {Mart\'in-Mayor},
  \citenamefont {noz Sudupe}, \citenamefont {Parisi},\ and\ \citenamefont
  {Ruiz-Lorenzo}}]{Ballesteros1999}%
  \BibitemOpen
  \bibfield  {author} {\bibinfo {author} {\bibfnamefont {H.~G.}\ \bibnamefont
  {Ballesteros}}, \bibinfo {author} {\bibfnamefont {L.~A.}\ \bibnamefont
  {Fern\'andez}}, \bibinfo {author} {\bibfnamefont {V.}~\bibnamefont
  {Mart\'in-Mayor}}, \bibinfo {author} {\bibfnamefont {A.~M.}\ \bibnamefont
  {noz Sudupe}}, \bibinfo {author} {\bibfnamefont {G.}~\bibnamefont {Parisi}},
  \ and\ \bibinfo {author} {\bibfnamefont {J.~J.}\ \bibnamefont
  {Ruiz-Lorenzo}},\ }\href {http://stacks.iop.org/0305-4470/32/i=1/a=004}
  {\bibfield  {journal} {\bibinfo  {journal} {J. Phys. A: Math. Gen.}\ }\textbf
  {\bibinfo {volume} {32}},\ \bibinfo {pages} {1} (\bibinfo {year}
  {1999})}\BibitemShut {NoStop}%
\bibitem [{\citenamefont {Deng}\ and\ \citenamefont
  {Bl\"ote}(2005)}]{PhysRevE.72.016126}%
  \BibitemOpen
  \bibfield  {author} {\bibinfo {author} {\bibfnamefont {Y.}~\bibnamefont
  {Deng}}\ and\ \bibinfo {author} {\bibfnamefont {H.~W.~J.}\ \bibnamefont
  {Bl\"ote}},\ }\href {\doibase 10.1103/PhysRevE.72.016126} {\bibfield
  {journal} {\bibinfo  {journal} {Phys. Rev. E}\ }\textbf {\bibinfo {volume}
  {72}},\ \bibinfo {pages} {016126} (\bibinfo {year} {2005})}\BibitemShut
  {NoStop}%
\bibitem [{\citenamefont {Wang}\ \emph {et~al.}(2013)\citenamefont {Wang},
  \citenamefont {Zhou}, \citenamefont {Zhang}, \citenamefont {Garoni},\ and\
  \citenamefont {Deng}}]{PhysRevE.87.052107}%
  \BibitemOpen
  \bibfield  {author} {\bibinfo {author} {\bibfnamefont {J.}~\bibnamefont
  {Wang}}, \bibinfo {author} {\bibfnamefont {Z.}~\bibnamefont {Zhou}}, \bibinfo
  {author} {\bibfnamefont {W.}~\bibnamefont {Zhang}}, \bibinfo {author}
  {\bibfnamefont {T.~M.}\ \bibnamefont {Garoni}}, \ and\ \bibinfo {author}
  {\bibfnamefont {Y.}~\bibnamefont {Deng}},\ }\href {\doibase
  10.1103/PhysRevE.87.052107} {\bibfield  {journal} {\bibinfo  {journal} {Phys.
  Rev. E}\ }\textbf {\bibinfo {volume} {87}},\ \bibinfo {pages} {052107}
  (\bibinfo {year} {2013})}\BibitemShut {NoStop}%
\bibitem [{\citenamefont {Jan}\ and\ \citenamefont
  {Stauffer}(1998)}]{Naeem1998}%
  \BibitemOpen
  \bibfield  {author} {\bibinfo {author} {\bibfnamefont {N.}~\bibnamefont
  {Jan}}\ and\ \bibinfo {author} {\bibfnamefont {D.}~\bibnamefont {Stauffer}},\
  }\href {\doibase 10.1142/S0129183198000261} {\bibfield  {journal} {\bibinfo
  {journal} {Int. J. Mod. Phys. C}\ }\textbf {\bibinfo {volume} {9}},\ \bibinfo
  {pages} {341} (\bibinfo {year} {1998})}\BibitemShut {NoStop}%
\bibitem [{\citenamefont {Acharyya}\ and\ \citenamefont
  {Stauffer}(1998)}]{Acharyya1998}%
  \BibitemOpen
  \bibfield  {author} {\bibinfo {author} {\bibfnamefont {M.}~\bibnamefont
  {Acharyya}}\ and\ \bibinfo {author} {\bibfnamefont {D.}~\bibnamefont
  {Stauffer}},\ }\href {\doibase 10.1142/S0129183198000534} {\bibfield
  {journal} {\bibinfo  {journal} {Int. J. Mod. Phys. C}\ }\textbf {\bibinfo
  {volume} {9}},\ \bibinfo {pages} {643} (\bibinfo {year} {1998})}\BibitemShut
  {NoStop}%
\bibitem [{\citenamefont {Grassberger}(1992)}]{Grassberger1992}%
  \BibitemOpen
  \bibfield  {author} {\bibinfo {author} {\bibfnamefont {P.}~\bibnamefont
  {Grassberger}},\ }\href {http://stacks.iop.org/0305-4470/25/i=22/a=015}
  {\bibfield  {journal} {\bibinfo  {journal} {J. Phys. A: Math. Gen.}\ }\textbf
  {\bibinfo {volume} {25}},\ \bibinfo {pages} {5867} (\bibinfo {year}
  {1992})}\BibitemShut {NoStop}%
\bibitem [{\citenamefont {Birenbaum}\ and\ \citenamefont
  {Ederer}(2014)}]{Birenbaum}%
  \BibitemOpen
  \bibfield  {author} {\bibinfo {author} {\bibfnamefont {A.~Y.}\ \bibnamefont
  {Birenbaum}}\ and\ \bibinfo {author} {\bibfnamefont {C.}~\bibnamefont
  {Ederer}},\ }\href {\doibase 10.1103/PhysRevB.90.214109} {\bibfield
  {journal} {\bibinfo  {journal} {Phys. Rev. B}\ }\textbf {\bibinfo {volume}
  {90}},\ \bibinfo {pages} {214109} (\bibinfo {year} {2014})}\BibitemShut
  {NoStop}%
\bibitem [{\citenamefont {Vandermeer}\ \emph {et~al.}(2010)\citenamefont
  {Vandermeer}, \citenamefont {Perfecto},\ and\ \citenamefont
  {Schellhorn}}]{Vandermeer}%
  \BibitemOpen
  \bibfield  {author} {\bibinfo {author} {\bibfnamefont {J.}~\bibnamefont
  {Vandermeer}}, \bibinfo {author} {\bibfnamefont {I.}~\bibnamefont
  {Perfecto}}, \ and\ \bibinfo {author} {\bibfnamefont {N.}~\bibnamefont
  {Schellhorn}},\ }\href {\doibase 10.1007/s10980-010-9449-2} {\bibfield
  {journal} {\bibinfo  {journal} {Landscape Ecol.}\ }\textbf {\bibinfo {volume}
  {25}},\ \bibinfo {pages} {509} (\bibinfo {year} {2010})}\BibitemShut
  {NoStop}%
\bibitem [{\citenamefont {Chen}\ and\ \citenamefont {Schuh}(2009)}]{Chen}%
  \BibitemOpen
  \bibfield  {author} {\bibinfo {author} {\bibfnamefont {Y.}~\bibnamefont
  {Chen}}\ and\ \bibinfo {author} {\bibfnamefont {C.~A.}\ \bibnamefont
  {Schuh}},\ }\href {\doibase 10.1103/PhysRevE.80.040103} {\bibfield  {journal}
  {\bibinfo  {journal} {Phys. Rev. E}\ }\textbf {\bibinfo {volume} {80}},\
  \bibinfo {pages} {040103} (\bibinfo {year} {2009})}\BibitemShut {NoStop}%
\bibitem [{\citenamefont {Bianconi}(2013)}]{Bianconi}%
  \BibitemOpen
  \bibfield  {author} {\bibinfo {author} {\bibfnamefont {G.}~\bibnamefont
  {Bianconi}},\ }\href {\doibase 10.1209/0295-5075/101/26003} {\bibfield
  {journal} {\bibinfo  {journal} {EPL}\ }\textbf {\bibinfo {volume} {101}},\
  \bibinfo {pages} {26003} (\bibinfo {year} {2013})}\BibitemShut {NoStop}%
\bibitem [{\citenamefont {Yiu}\ \emph {et~al.}(2014)\citenamefont {Yiu},
  \citenamefont {Bonf\'a}, \citenamefont {Sanna}, \citenamefont {De~Renzi},
  \citenamefont {Carretta}, \citenamefont {McGuire}, \citenamefont {Huq},\ and\
  \citenamefont {Nagler}}]{Yiu}%
  \BibitemOpen
  \bibfield  {author} {\bibinfo {author} {\bibfnamefont {Y.}~\bibnamefont
  {Yiu}}, \bibinfo {author} {\bibfnamefont {P.}~\bibnamefont {Bonf\'a}},
  \bibinfo {author} {\bibfnamefont {S.}~\bibnamefont {Sanna}}, \bibinfo
  {author} {\bibfnamefont {R.}~\bibnamefont {De~Renzi}}, \bibinfo {author}
  {\bibfnamefont {P.}~\bibnamefont {Carretta}}, \bibinfo {author}
  {\bibfnamefont {M.~A.}\ \bibnamefont {McGuire}}, \bibinfo {author}
  {\bibfnamefont {A.}~\bibnamefont {Huq}}, \ and\ \bibinfo {author}
  {\bibfnamefont {S.~E.}\ \bibnamefont {Nagler}},\ }\href {\doibase
  10.1103/PhysRevB.90.064515} {\bibfield  {journal} {\bibinfo  {journal} {Phys.
  Rev. B}\ }\textbf {\bibinfo {volume} {90}},\ \bibinfo {pages} {064515}
  (\bibinfo {year} {2014})}\BibitemShut {NoStop}%
\bibitem [{\citenamefont {Smits}\ \emph {et~al.}(1993)\citenamefont {Smits},
  \citenamefont {Koper},\ and\ \citenamefont {Mandel}}]{Smits}%
  \BibitemOpen
  \bibfield  {author} {\bibinfo {author} {\bibfnamefont {R.~G.}\ \bibnamefont
  {Smits}}, \bibinfo {author} {\bibfnamefont {G.~J.~M.}\ \bibnamefont {Koper}},
  \ and\ \bibinfo {author} {\bibfnamefont {M.}~\bibnamefont {Mandel}},\ }\href
  {\doibase 10.1021/j100123a047} {\bibfield  {journal} {\bibinfo  {journal} {J.
  Phys. Chem.}\ }\textbf {\bibinfo {volume} {97}},\ \bibinfo {pages} {5745}
  (\bibinfo {year} {1993})}\BibitemShut {NoStop}%
\bibitem [{\citenamefont {Wang}\ \emph {et~al.}(2014)\citenamefont {Wang},
  \citenamefont {Zhang},\ and\ \citenamefont {Zhang}}]{Wang}%
  \BibitemOpen
  \bibfield  {author} {\bibinfo {author} {\bibfnamefont {R.}~\bibnamefont
  {Wang}}, \bibinfo {author} {\bibfnamefont {Y.}~\bibnamefont {Zhang}}, \ and\
  \bibinfo {author} {\bibfnamefont {C.}~\bibnamefont {Zhang}},\ }\href
  {\doibase 10.1016/j.rinp.2014.03.006} {\bibfield  {journal} {\bibinfo
  {journal} {Results in Physics}\ }\textbf {\bibinfo {volume} {4}},\ \bibinfo
  {pages} {44 } (\bibinfo {year} {2014})}\BibitemShut {NoStop}%
\bibitem [{\citenamefont {Kriel}\ \emph {et~al.}(2014)\citenamefont {Kriel},
  \citenamefont {Karrasch},\ and\ \citenamefont {Kehrein}}]{Kriel}%
  \BibitemOpen
  \bibfield  {author} {\bibinfo {author} {\bibfnamefont {J.~N.}\ \bibnamefont
  {Kriel}}, \bibinfo {author} {\bibfnamefont {C.}~\bibnamefont {Karrasch}}, \
  and\ \bibinfo {author} {\bibfnamefont {S.}~\bibnamefont {Kehrein}},\ }\href
  {\doibase 10.1103/PhysRevB.90.125106} {\bibfield  {journal} {\bibinfo
  {journal} {Phys. Rev. B}\ }\textbf {\bibinfo {volume} {90}},\ \bibinfo
  {pages} {125106} (\bibinfo {year} {2014})}\BibitemShut {NoStop}%
\bibitem [{\citenamefont {Rodrigues}\ and\ \citenamefont
  {Oliveira}(2014)}]{Rodrigues}%
  \BibitemOpen
  \bibfield  {author} {\bibinfo {author} {\bibfnamefont {N.~T.}\ \bibnamefont
  {Rodrigues}}\ and\ \bibinfo {author} {\bibfnamefont {T.~J.}\ \bibnamefont
  {Oliveira}},\ }\href {http://stacks.iop.org/1751-8121/47/i=40/a=405002}
  {\bibfield  {journal} {\bibinfo  {journal} {J. Phys. A: Math. Theor.}\
  }\textbf {\bibinfo {volume} {47}},\ \bibinfo {pages} {405002} (\bibinfo
  {year} {2014})}\BibitemShut {NoStop}%
\bibitem [{\citenamefont {Gratens}\ \emph {et~al.}(2007)\citenamefont
  {Gratens}, \citenamefont {Paduan-Filho}, \citenamefont {Bindilatti},
  \citenamefont {Oliveira},\ and\ \citenamefont {Shapira}}]{Gratens}%
  \BibitemOpen
  \bibfield  {author} {\bibinfo {author} {\bibfnamefont {X.}~\bibnamefont
  {Gratens}}, \bibinfo {author} {\bibfnamefont {A.}~\bibnamefont
  {Paduan-Filho}}, \bibinfo {author} {\bibfnamefont {V.}~\bibnamefont
  {Bindilatti}}, \bibinfo {author} {\bibfnamefont {N.~F.}\ \bibnamefont
  {Oliveira}, \bibfnamefont {Jr.}}, \ and\ \bibinfo {author} {\bibfnamefont
  {Y.}~\bibnamefont {Shapira}},\ }\href {\doibase 10.1103/PhysRevB.75.184405}
  {\bibfield  {journal} {\bibinfo  {journal} {Phys. Rev. B}\ }\textbf {\bibinfo
  {volume} {75}},\ \bibinfo {pages} {184405} (\bibinfo {year}
  {2007})}\BibitemShut {NoStop}%
\bibitem [{\citenamefont {Albert}\ \emph {et~al.}(2008)\citenamefont {Albert},
  \citenamefont {Schnabel},\ and\ \citenamefont {Field}}]{Albert}%
  \BibitemOpen
  \bibfield  {author} {\bibinfo {author} {\bibfnamefont {M.~V.}\ \bibnamefont
  {Albert}}, \bibinfo {author} {\bibfnamefont {A.}~\bibnamefont {Schnabel}}, \
  and\ \bibinfo {author} {\bibfnamefont {D.~J.}\ \bibnamefont {Field}},\ }\href
  {\doibase 10.1371/journal.pcbi.1000137} {\bibfield  {journal} {\bibinfo
  {journal} {{PLoS} Comput. Biol.}\ }\textbf {\bibinfo {volume} {4}},\ \bibinfo
  {pages} {e000137} (\bibinfo {year} {2008})}\BibitemShut {NoStop}%
\bibitem [{\citenamefont {Boer}\ \emph {et~al.}(2011)\citenamefont {Boer},
  \citenamefont {Johnston},\ and\ \citenamefont {Sadler}}]{Boer}%
  \BibitemOpen
  \bibfield  {author} {\bibinfo {author} {\bibfnamefont {M.~M.}\ \bibnamefont
  {Boer}}, \bibinfo {author} {\bibfnamefont {P.}~\bibnamefont {Johnston}}, \
  and\ \bibinfo {author} {\bibfnamefont {R.~J.}\ \bibnamefont {Sadler}},\
  }\href {\doibase 10.1016/j.ecocom.2011.07.005} {\bibfield  {journal}
  {\bibinfo  {journal} {Ecol. Complex.}\ }\textbf {\bibinfo {volume} {8}},\
  \bibinfo {pages} {347} (\bibinfo {year} {2011})}\BibitemShut {NoStop}%
\end{thebibliography}%

%% ############################################################################
\end{document}